# Generalized echo squeezing protocol with near-Heisenberg-limit sensitivity and strong robustness against detection noise and variation in squeezing parameter

Jinyang Li[1], Gregório R. M. da Silva[1], Schuyler Kain[1], Selim M. Shahriar[2,3]
[1] Department of Physics and Astronomy, Northwestern University, Evanston, IL 60208, USA
[2] Department of ECE, Northwestern University, Evanston, IL 60208, USA
[3] Digital Optics Technologies, Rolling Meadows, IL 60008, USA

## Abstract

We present a generalized echo squeezing protocol (GESP) as a generalization of the Schrödinger cat state protocol (SCSP) with the value of the squeezing parameter being an arbitrary number rather than $\pi/2$. We show analytically that over a very broad range of the squeezing parameter the sensitivity of the GESP reaches the Heisenberg limit within a factor of $\sqrt{2}$. For a large number of atoms, $N$, this plateau interval is almost the whole range from zero to $\pi/2$, and the sensitivity is independent of the parity of $N$. Therefore, it is possible to operate a sensor over a wide interval of the squeezing parameter without changing the sensitivity. This is to be contrasted with the conventional echo squeezing protocol (CESP) which only works for a very small interval of the squeezing parameter. We also show that, in contrast to the CESP, the sensitivity of the GESP is close to the quantum Cramér-Rao bound over the whole range of the squeezing parameter, indicating that the phase shift information contained in the quantum state is near-optimally extracted. We find that the enhancement in sensitivity in the case of the GESP is due to a combination of two parameters: the phase magnification factor (PMF) and the noise amplification factor (NAF). As the value of the squeezing parameter increases, both PMF and NAF increase, while keeping the ratio of PMF/NAF essentially constant, yielding a net enhancement of sensitivity at the Heisenberg limit within a factor of $\sqrt{2}$ over the whole plateau interval. An important consequence of this behavior is that the robustness of the GESP against detection noise easily exceeds that of the CESP for a broad range of values of the squeezing parameter. As such, in the context of an experimental study, it should be possible to achieve a net enhancement in sensitivity higher than that for the CESP, under typical conditions where detection noise exceeds the quantum projection noise of an unsqueezed state with the same number of atoms. Finally, we consider the fragility of the GESP against decoherence mechanisms, and show how a balance between the fragility against the decoherence mechanisms and the robustness against detection noise would in practice determine the optimal choice of parameters for the GESP.

## 1. Introduction

The conventional echo squeezing protocol (CESP) [1,2] is a method of enhancing the sensitivity of an atomic sensor using one-axis-twist spin squeezing (OATS) [3,4,5,6,7,8,9]. The Hamiltonian for OATS can be expressed as $\chi S_z^2$, where $\chi$ represents the strength of the non-linear interaction, and $S_z$ is the sum of the *z*-components of the pseudo-spin vector for each atom, represented as a two level system. The effect of the OATS process on the collective response of the atoms is determined by the squeezing parameter, define as $\mu \equiv \chi t$, where $t$ is the squeezing interaction time. The CESP can be applied to many atomic sensors, including an atomic clock and a light-pulse atomic interferometer. For concreteness, we first restrict the discussion to the case of the Ramsey atomic clock. The case of the atom interferometer will be discussed in detail in Section 2. To recall briefly, the CESP-enhanced Ramsey clock works as follows [10]. We presume that the atoms are first prepared in the state where all the pseudo-spinors are in the *z*-direction. We further assume that the first π/2 pulse causes a rotation around the *y*-axis, thus making all pseudo-spins point in the *x*-direction. This is followed by the OATS process, applied for an optimal value of $\mu \approx N^{-1/2}$, where $N$ is the number of atoms being interrogated. Next, an auxiliary rotation, by an angle of π/2, is carried out around the *x*-axis, before the evolution during the dark period. At the end of the dark period, the auxiliary rotation is reversed, and the OATS process is applied again, with the sign of the squeezing Hamiltonian reversed (hence the use of the word "echo"). This is followed by another π/2 pulse, but around the *x*-axis, rather than the *y*-axis. The signal measured is the expectation value of $S_z$. It has been shown [1] that, under ideal conditions, the sensitivity is enhanced by a factor of $\sim \sqrt{N/\mathrm{e}}$. It has also been shown [1,2] that compared to protocols without spin squeezing, this enhancement is due only to a phase magnification, without any variation in the quantum projection noise. As a result, the actual factor of enhancement in sensitivity achievable is significantly less than the ideal value if detection noise is far beyond the quantum projection noise of an unsqueezed ensemble also with $N$ atoms [1].

Recently, we had proposed an alternative protocol [11] using OATS that also employs inversion of the squeezing interaction. For this protocol, we had focused on the case of $\mu = \pi/2$, which produces the Schrödinger cat state, yielding the Heisenberg limit sensitivity if the parity of $N$ is known. As such, we will refer to this as the Schrödinger cat state protocol (SCSP). In addition to the Heisenberg limit sensitivity, the phase magnification and the quantum projection noise of the SCSP are also maximized among all the protocols using OATS. Considering that the robustness can be defined as the level of detection noise that will reduce the sensitivity by a factor of $\sqrt{2}$, we can easily see that the robustness equals the level of the quantum projection noise. Therefore, the SCSP is the most robust to detection noise among all protocols using OATS [12]. However, the SCSP is extremely fragile to decoherence mechanisms such as cavity decay, residual spontaneous emission, and collisions between atoms and background particles, making it less promising. Therefore, it is important to investigate this protocol for a smaller value of $\mu$ for which the protocol is less fragile. The sensitivity for the whole range of the squeezing parameter $0 \le \mu \le \pi/2$ computed with numerical simulations is also shown in Ref. [11]. However, the cursory study of this protocol for $0 \le \mu \le \pi/2$ shown in Ref. [11] is inadequate for judging whether this protocol for a value of $\mu$ less than π/2 is a better alternative to the CESP and the SCSP for three reasons. First, only the sensitivity was numerically computed; the robustness to detection noise and the fragility to decoherence mechanisms were not studied. Second, in numerical simulations, exponential functions of non-diagonal $(N+1)\times(N+1)$ matrices need to be calculated, which takes an unacceptable long time for ordinary computation resources if the value of $N$ exceeds a few hundred, far less than the value of $N$ expected in a real experiment. Third, no universal conclusion can be drawn solely with numerical simulations implemented for a few values of $N$. In this paper, we present a detailed and analytical study of this protocol for the range $0 \leq \mu \leq \pi/2$. Specifically, we derive the analytical expressions for the phase magnification, the noise amplification, and thereby the sensitivity, for any value of $\mu$ and any value of $N$. In addition, we determine how the robustness of the protocol against detection noise varies as a function of the squeezing parameter. Furthermore, we show how the fragility of this state against collision with background particles depend on the value of the squeezing parameter. Given that this protocol also employs the unsqueezing process, just like the CESP, we choose to call it the generalized echo squeezing protocol (GESP). We note that since the steps in the GESP are exactly the same as those for the SCSP, the results of our analysis for the GESP, as presented in this paper, will in some cases include the SCSP as the limiting case.

The basic steps of the GESP are the same as the ones described above for the CESP. The differences are only the values of $\mu$ the protocols should be operated at and the rotation axes of the $\pi/2$ pulses (including the auxiliary rotation and its inverse). Perhaps surprisingly, these two differences produce drastically different results. To explicitly show the differences, we note first that the range of $\mu$ of interest for OATS is from 0 to π/2 because the behavior of the system for values of $\mu$ from π/2 to π mirrors that for values of $\mu$ from 0 to π/2, and has an overall periodicity of π. Next, it should be noted that while the results of the CESP protocol does not depend on the parity of $N$, the results of the GESP protocol depend strongly on this parity for the values of $\mu$ close to π/2. Specifically, this dependence on the parity of $N$ occurs in the interval $\left(\pi/2-\sqrt{2/N}\right) \leq \mu \leq \pi/2$, as shown in Section 3. Therefore, in order to understand the dynamics of the GESP fully, including its behavior in the bridge region, it is necessary to describe two versions of the GESP: one optimized for odd $N$ in the interval $\left(\pi/2-\sqrt{2/N}\right)\leq\mu\leq\pi/2$, and the other optimized for even $N$ in that interval.

We denote as GESP-o (GESP-e) the version of the GESP optimized for odd (even) $N$. For GESP-o, all the steps are the same as those for the CESP described above, with the exception that the last π/2 pulse rotation is around the *y*-axis. For GESP-e, the steps are similar to those used for the CESP, with the following differences. First, the auxiliary rotation after the OATS is carried out around the *y*-axis, and the reversal of the auxiliary rotation is also around the *y*-axis. Second, the last π/2 pulse rotation is around the *y*-axis. For both versions of the GESP, the signal measured is the expectation value of $S_z$, just as in the case of the CESP. The case investigated in Ref. [11] was focused on optimizing the SCSP for even values of *N*. As such, when considering the whole range of the squeezing parameter, it would correspond to GESP-e.

Our detailed and analytical study on the GESP shows the following results. In the interval $0\leq\mu\leq1/\sqrt{N}$, the sensitivity of the GESP (for both versions) is close to the sensitivity of the CESP, and even exceeds that sensitivity for some values of $\mu$. As $\mu$ increases from $1/\sqrt{N}$, the sensitivity of the CESP decreases to zero rapidly while that of the GESP (for both versions) keeps increasing, reaching the Heisenberg limit, within a factor of $\sqrt{2}$, at about $\mu=4\sqrt{2/N}$, and remains almost constant until $\mu=\left(\pi/2-\sqrt{2/N}\right)$. As $\mu$ increase from $\left(\pi/2-\sqrt{2/N}\right)$ to $\pi/2$, the sensitivity of the GESP goes to either the Heisenberg limit or the standard quantum limit, depending on the combination of the version of the GESP and the parity of $N$. We also compare the sensitivity of the GESP (including the SCSP as a limiting case) and the CESP with the corresponding quantum Cramér-Rao bounds (QCR bounds). We find that for the GESP (both versions) the sensitivity remains very close to the QCR bound for the whole range of the squeezing parameter, while for the CESP the sensitivity is far below the QCR bound for values of $\mu$ beyond the optimal one. This result indicates that the phase shift information contained in the quantum state is near-optimally extracted in the case of the GESP.

We also compare the mechanisms of enhancement of sensitivity for the GESP (including the limits case of the SCSP) and the CESP. In the case of the SCSP, the enhancement of sensitivity is due to an $N$-fold phase magnification, countered by a $\sqrt{N}$ -fold magnification of the quantum projection noise compared to an unsqeezed ensemble also with $N$ atoms. In contrast, for the CESP, at its optimum point of operation, the enhancement is due only to a phase magnification by a factor of $\sim\sqrt{N/\mathrm{e}}$, with no variation in the quantum projection noise compared to an unsqueezed ensemble. For the case of GESP, the mechanism depends on the value of $\mu$. Consider specifically the range of $\mu$ over which the enhancement of sensitivity remains essentially a constant, with a value of $\sim\sqrt{N/2}$. For small values of $\mu$, the enhancement mechanism is similar to that of the CESP: it is due primarily to phase magnification by a factor of $\sim\sqrt{N/2}$, with no change in the quantum projection noise. For the values of $\mu$ close to $\pi/2$, the enhancement mechanism is similar to that of the SCSP: there is a phase magnification by a factor of $\sim N/\sqrt{2}$, countered by an increase in the quantum projection noise by a factor of $\sim\sqrt{N}$. We note that the details of the mechanism of enhancement, while seemingly irrelevant, does in fact have great significance in the experimental context. As we showed in detail in Ref. [11], the extreme phase magnification and the corresponding amplification of the quantum noise in the case of SCSP makes it very robust against detection noise. In fact, as mentioned earlier, the protocol presented in Ref. [11] has been shown to achieve the maximum possible robustness against detection noise [12]. When the GESP is operated at the point where $\mu$ is close to π/2, it has a robustness against detection noise almost as high as in the case of the SCSP. We also determine the shapes of the interference fringes for the GESP and the CESP to prove the phase magnification effect discussed above and predict the interference fringes that would be observed when the phase shift is scanned.

Despite the high robustness against detection noise as $\mu$ approaches π/2 for the GESP, the fragility against decoherence mechanisms also increases. In this paper, we consider the decoherence mechanisms including cavity decay, residual spontaneous emission, and collisions with background particles, for both GESP (including the limiting case of the SCSP) as well as the CESP. We find that the fragility of both the protocols, expressed in terms of the reduction in the visibility of the signal fringes as a function of the phase shift, increases monotonically with $\mu$. However, the CESP can only achieve a near Heisenberg limit sensitivity in the proximity of the optimal value of $\mu$ (i.e., $\mu \approx 1/\sqrt{N}$), making it meaningless to discuss its fragility far from the point $\mu \approx 1/\sqrt{N}$. The wider choice of the value of $\mu$ makes the GESP a very attractive option experimentally.

The discussions presented above clearly show that the GESP is potentially a better alternative to the SCSP and the CESP experimentally. To summarize, for the GESP it is not necessary to tune the value of the squeezing parameter to a sharply peaked value (for the SCSP, this peak value is $\mu = \pi/2$ and for the CESP $\mu \approx 1/\sqrt{N}$). Furthermore, the GESP is less vulnerable to detection noise than the CESP and less fragile to decoherence mechanisms than the SCSP. In addition, the GESP can be implemented for an extremely broad range of values of $\mu$ with the same (near

Heisenberg limit) sensitivity, providing a widely adjustable range so that one can balance the requirements for the robustness to detection noise and the resistance to decoherence mechanisms.

The rest of the paper is organized as follows. We describe the GESP together with the SCSP and the CESP in Section 2, calculate the sensitivity of the GESP in Section 3, investigate the behavior of the sensitivity in Section 4, analyze the mechanism of the enhancement of sensitivity to show the robustness of the GESP against detection noise in Section 5, analyze the fragility of all the protocols against decoherence mechanisms in Section 6, and present concluding remarks in Section 7.

## 2. Generalized echo squeezing protocol

We first review briefly the canonical two-level atom in order to establish the relevant notations used in this paper. A two-level atom is formally equivalent to a spin-1/2 spinor, with the spin operator represented as $\boldsymbol{s} = \left(s_x, s_y, s_z\right)$, with $s_w \left(w = x, y, z\right)$ being the component of the spin operator in the $w$-direction. The two eigenstates of $s_w$ are denoted as $\left|\hat{\boldsymbol{w}}_0\right\rangle$ and $\left|-\hat{\boldsymbol{w}}_0\right\rangle$ (e.g., $\left|\hat{\boldsymbol{z}}_0\right\rangle$ and $\left|-\hat{\boldsymbol{z}}_0\right\rangle$) with eigenvalues of $1/2$ and $-1/2$, respectively (setting $\hbar = 1$). For an ensembles of $N$ non-interacting atoms, the combined spin-operator is defined as:

$$\boldsymbol{S} \equiv \left(S_x, S_y, S_z\right) \equiv \sum_{j=1}^{N} \boldsymbol{s}_j \tag{1}$$

The coherent spin state (CSS) with all the atoms oriented in the $\pm w$ direction is denoted as $\left|\pm\hat{\boldsymbol{w}}\right\rangle$.

To describe the generalized echo squeezing protocol (GESP), we start by considering the limiting case, corresponding to the SCSP, where the squeezing parameter, $\mu$, has the value of $\pi/2$. We recall [11] that there are two types of SCSP: one optimized for even $N$ (denoted as SCSP-e), while the other optimized for odd $N$ (denoted as SCSP-o). The SCSP-e (SCSP-o) reaches the Heisenberg limit of sensitivity when $N$ is even (odd) [13,14,15,16,17,18]. The steps in the SCSP-e, applied to a Ramsey atomic clock are illustrated schematically in Figure 1, using the Husimi quasi-probability distribution [3] for both even (top row) and odd (bottom row) values of $N$. First, a π/2 pulse around the *y*-axis is applied to the atoms in state $\left|\hat{\boldsymbol{z}}\right\rangle$, which turns the atoms into CSS $\left|\hat{\boldsymbol{x}}\right\rangle$. The effect of this step is equivalent to starting with atoms in CSS $\left|\hat{\boldsymbol{x}}\right\rangle$. Then the atoms go through an OATS process with $\mu=\pi/2$, which produces a Schrödinger cat state oriented along the *x*-axis. Since the phase shift due to the detuning of the clock oscillator, denoted as $\phi$, causes a rotation around the *z*-axis, an auxiliary pulse is applied to orient the Schrödinger cat state along the *z*-axis. This auxiliary rotation is then undone at the end of the dark period. As we show later in the paper, the combined processes of applying this auxiliary rotation, and the accumulation of the phase shift around the *z*-axis, followed by reserving the auxiliary rotation, is equivalent to a rotation around the *x*-axis by the same angle $\phi$. As such, for simplicity and brevity, we represent these three steps as a single step entailing a rotation around the *x*-axis. Next, the atoms go through an inverse OATS process with $\mu=-\pi/2$. This is followed by another π/2 pulse around the *y*-axis, and the quantum operator that is measured is $S_z$. However, it can be shown that the combined effect of these two steps is equivalent to measuring the operator $S_x$. As such, for simplicity and brevity, we represent these two steps as a single one representing measurement of $S_x$. As can be seen by comparing the top row with the bottom one, the evolution of the quantum states during various stages of the protocol for even *N* differ significantly from those generated for odd *N*. This difference between the two parities of $N$ is due to the fact that the Schrödinger cat state generated by the first OATS process is *x*-directed (top row) for even $N$ but *y*-directed (bottom row) for odd $N$ . An explanation of why the addition of a single atom changes the orientation of the Schrödinger cat state by $\pi/2$ is given in Appendix A.

In Figure 2, we illustrate the corresponding steps for the SCSP-o, with the top (bottom) row corresponding to even (odd) values of *N*. The first two steps are the same as those for the SCSP-e. However, since the Schrödinger cat generated after the application of the OATS pulse is oriented around the *y*-axis, the auxiliary rotation used for aligning them along the *z*-axis and inversion thereof at the end of the dark zone occurs around the *x*-axis. As we also show later in this paper, the combined effect of the processes of the auxiliary rotation, the rotation around the *z*-axis by the angle $\phi$, and reversing the auxiliary rotation, is a rotation around the *y*-axis by the same angle $\phi$, as indicated here. The remainder of the protocol is identical to that of the SCSP-e. Therefore, we see that the SCSP-o differs from the SCSP-e only in the fact that the rotation representing the phase shift occurs around the *y*-axis (in practice, this represents a difference in the rotation axes of the auxiliary rotation and its inverse). Again, we see that the evolution of the quantum states depends significantly on the parity of *N*, for the same reason as the one described above for SCSP-e.

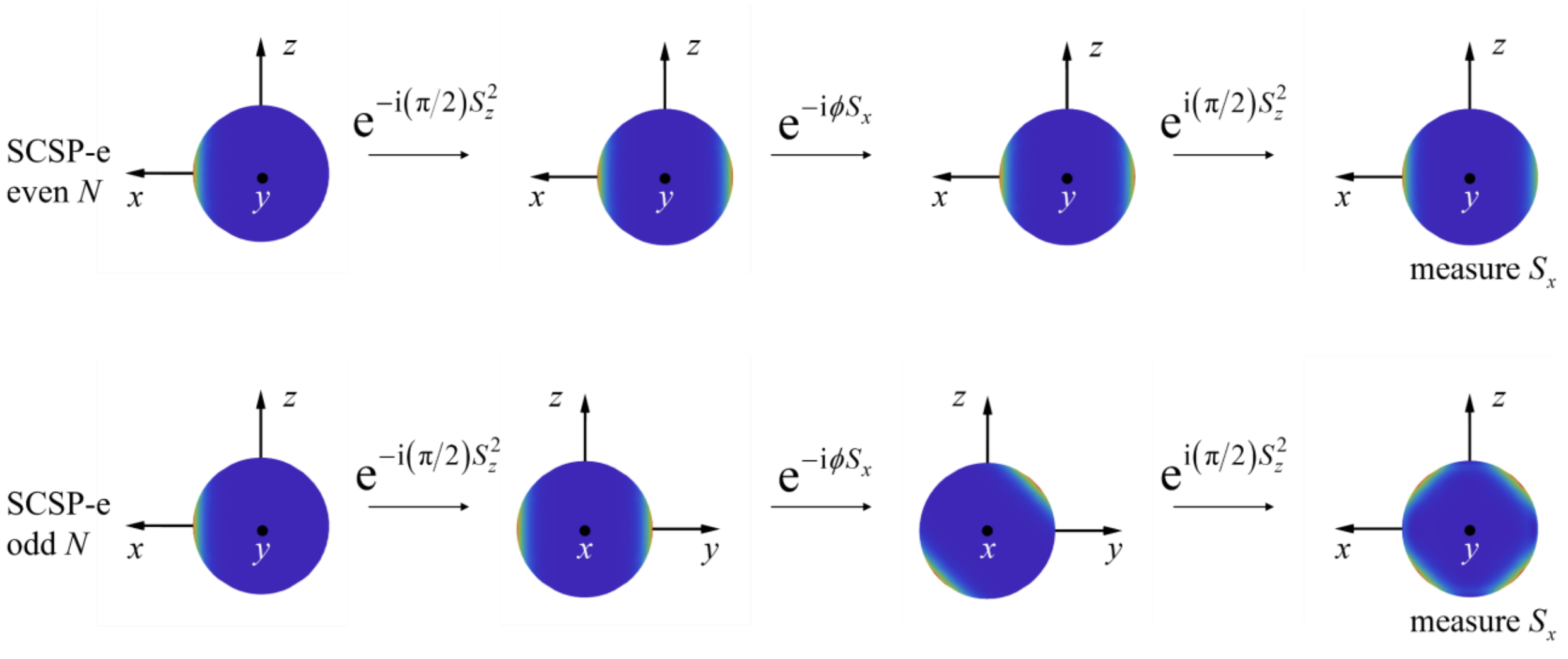

Figure 1. Steps of the Schrödinger cat state protocol optimized for even $N$ (SCSP-e) and the state evolution for even $N$ (top row) and odd $N$ (bottom row) illustrated by the Husimi quasi-probability distributions. The atoms prepared in the CSS $|\hat{x}\rangle$ first go through the OATS process with $\mu = \pi/2$, then accumulate phase shift due to the clock detuning, denoted as $\phi$. Here, we have made use of the fact that the effective rotation due the phase shift occurs around the $x$-axis, as explained in detail in the body of the paper. Finally, the atoms go through an inverse OATS process with $\mu = -\pi/2$. The effective quantum operator we measure in the end is $S_x$, as also explained in the body of the paper. As can be seen by comparing the top row with the bottom one, the quantum states generated for even $N$ differ significantly from those generated for odd $N$.

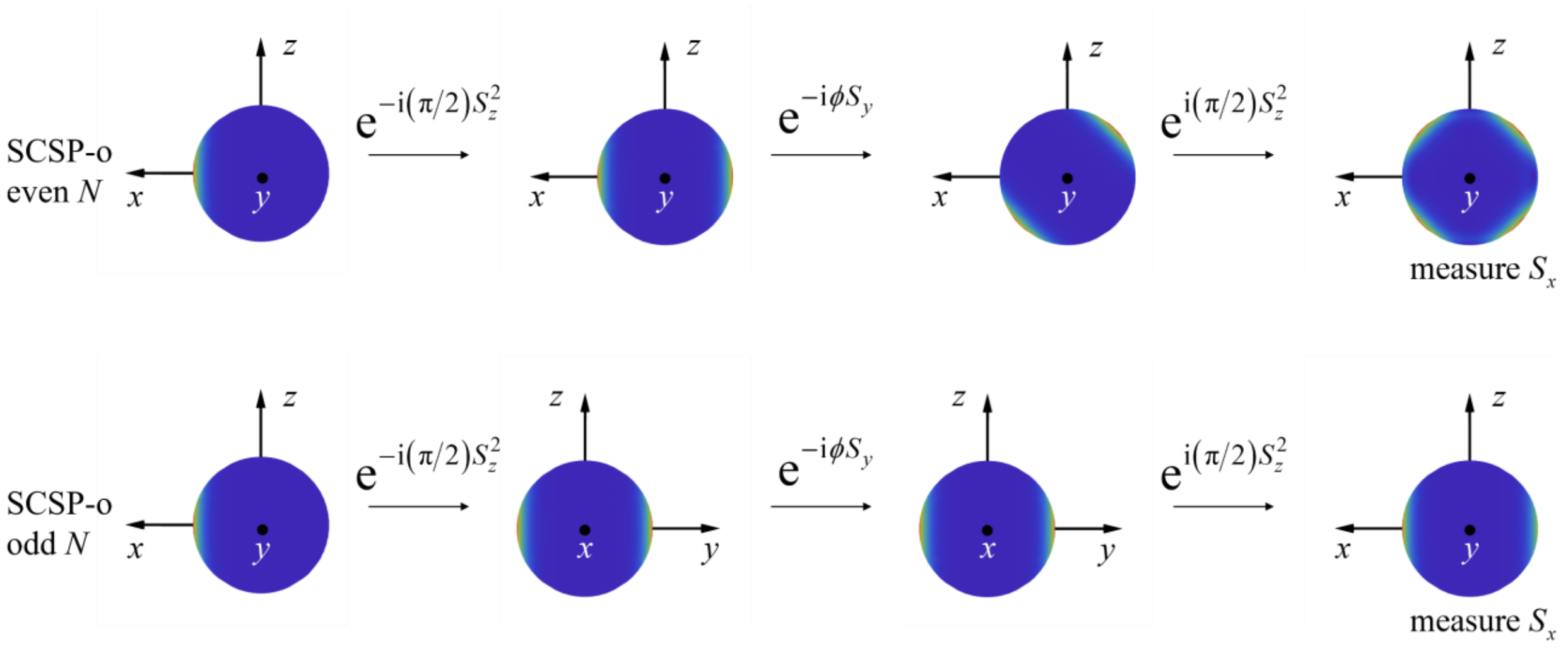

Figure 2. Steps of the Schrödinger cat state protocol optimized for odd $N$ (SCSP-o) and the state evolution for even $N$ (top row) and odd $N$ (bottom row) illustrated by the Husimi quasi-probability distributions. It differs from the SCSP-e process (as illustrated in Figure 1) only in the fact that the effective phase shift induced rotation occurs around the $y$-axis. Again, as can be seen by comparing the top row with the bottom one, the quantum states generated for odd $N$ differ significantly from those generated for even $N$.

The two types of the GESP, namely GESP-e and GESP-o, are generalized versions of SCSP-e and SCSP-o described above, for an arbitrary value of the squeezing parameter, $\mu$. In the top two rows of Figure 3, we illustrate the behavior of the GESP-e and GESP-o, respectively, for $N = 40$ and $\mu = \text{arccot}\sqrt{N-2} \approx 1/\sqrt{N}$. This value of $\mu$ is chosen for direct comparison with the CESP, because it gives the optimal enhancement of sensitivity for the CESP [1]. To compare the behavior of the quantum states under the two versions of the GESP with that of the CESP, this value of $\mu$ is used for all these protocols. As we will show later, the behavior of the quantum states under either version of the GESP generally does *not* depend on the parity of *N*, until the value of $\mu$ gets very close to π/2, and this condition is satisfied for the plots shown in the top two rows. In the bottom row of Figure 3, we illustrate the steps employed in the CESP. Here, we have again used the approach where an equivalent simple rotation around an axis is used to represent the combined effect of multiple steps that would be required experimentally. In order to see clearly the similarities and difference between the CESP and the two versions of the GESP, it is instructive to look instead at the actual steps necessary to realize the CESP experimentally for an atomic clock, as illustrated in Figure 4, for the value of $\mu$ that yields the optimal enhancement in sensitivity. Just as in the case of each version of the GESP protocol, the process starts with the application of a π/2 pulse around the *y*-axis to the atoms in state $|\hat{\boldsymbol{z}}\rangle$, to produce the CSS $|\hat{\boldsymbol{x}}\rangle$. The effect of this step is equivalent to starting with the atoms in the CSS $|\hat{\boldsymbol{x}}\rangle$. The atoms then go through an OATS process with $\mu = \text{arccot}\sqrt{N-2} \approx 1/\sqrt{N}$. This is followed in sequence by three steps: an auxiliary rotation around the *x*-axis, the rotation around the *z*-axis by an angle of $\phi$ due to the phase shift, and the reversal of the auxiliary rotation. These three steps are equivalent to a rotation around the *y*-axis by an angle of $\phi$ (as shown in the bottom row of Figure 3), according to the relation:

$$R_{\substack{x\\y}}(\phi)=R_{\substack{y\\x}}\left(\pm\frac{\pi}{2}\right)R_z(\phi)R_{\substack{y\\x}}\left(\mp\frac{\pi}{2}\right) \tag{2}$$

where, for example, $R_z(\phi)\equiv e^{-i\phi S_z}$ is the rotation operator. Next, the atoms go through an inverse OATS process. The net effect of these additional four steps (namely, squeezing, auxiliary rotation, inversion of auxiliary rotation, and inverse squeezing) is approximately a rotation of the CSS $|\hat{\boldsymbol{x}}\rangle$ around the *z*-axis by an angle of $\sim\phi\sqrt{N/\mathrm{e}}$ (for small $\phi$). As such, the net effect of the process is a phase magnification by a factor of $\sim\sqrt{N/\mathrm{e}}$ [19]. To measure this rotation around the *z*-axis, the CESP protocol uses a π/2 pulse around the *x*-axis, followed by a measurement of $S_z$ [20]. Measuring $S_z$ after the last π/2 pulse around the *x*-axis is equivalent to measuring $S_y$ (as shown in the bottom row of Figure 3), according to the relation:

$$\langle\psi|S_{\substack{x\\y}}|\psi\rangle=\langle\psi|R_{\substack{y\\x}}\left(\pm\frac{\pi}{2}\right)S_zR_{\substack{y\\x}}\left(\mp\frac{\pi}{2}\right)|\psi\rangle \tag{3}$$

It should be noted that the rotation axis of this last π/2 pulse is different from that of the first π/2 pulse, in contrast to the conventional Ramsey protocol, the SCSP, and the GESP.

We now focus on the equivalent and simplified representations of the protocols as summarized in Figure 3. As can be seen, the CESP differs significantly from the GESP-e, but is very similar to the GESP-o. They are identical except only the quantum operator we measure in the end. In the GESP-o, we measure $S_x$ while in the CESP we measure $S_y$. When considering the actual steps involved in these protocols, the difference between the GESP-o the CESP is the rotation axis of the last π/2 pulse before detection.

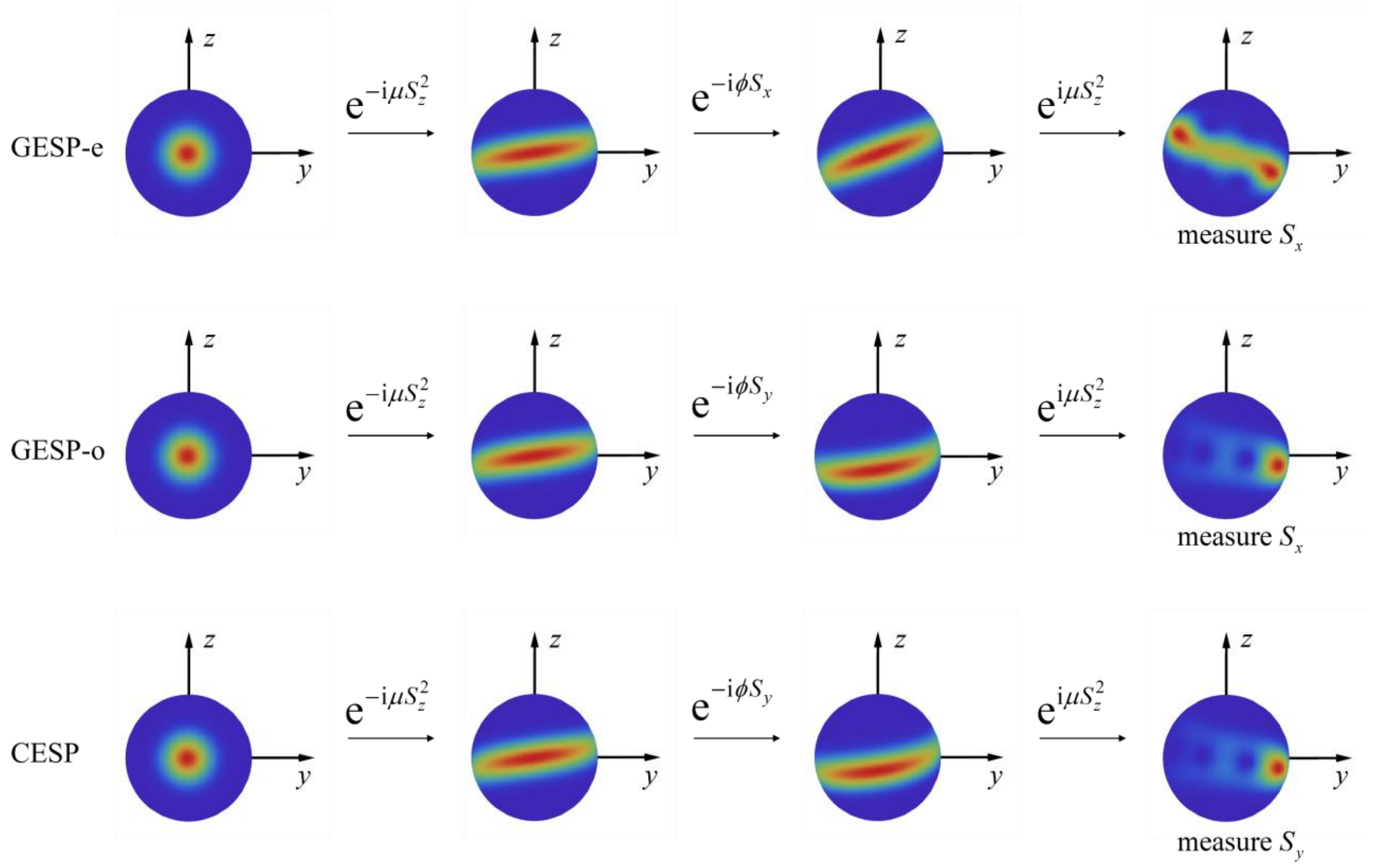

Figure 3. Generalized echo squeezing protocol GESP-e and GESP-o and the conventional echo squeezing protocol (CESP). In the top two rows, we illustrate the behavior of the GESP-e and GESP-o, respectively, for $N = 40$ and $\mu = \operatorname{arccot}\sqrt{N-2} \approx 1/\sqrt{N}$ . This value of $\mu$ is chosen for direct comparison with the CESP, because it gives the optimal enhancement of sensitivity for the CESP [1]. To compare the behavior of the quantum states under the two versions of the GESP with that of the CESP, this value of $\mu$ is used for all these protocols. As we will show later, the behavior of the quantum states under either version of the GESP generally does *not* depend on the parity of *N*, until the value of $\mu$ gets very close to π/2, and this condition is satisfied for the plots shown in the top two rows. The CESP differs significantly from the GESP-e, but is very similar to the GESP-o. They are identical except only the quantum operator we measure in the end. In the GESP-o, we measure $S_x$ while in the CESP we measure $S_y$ .

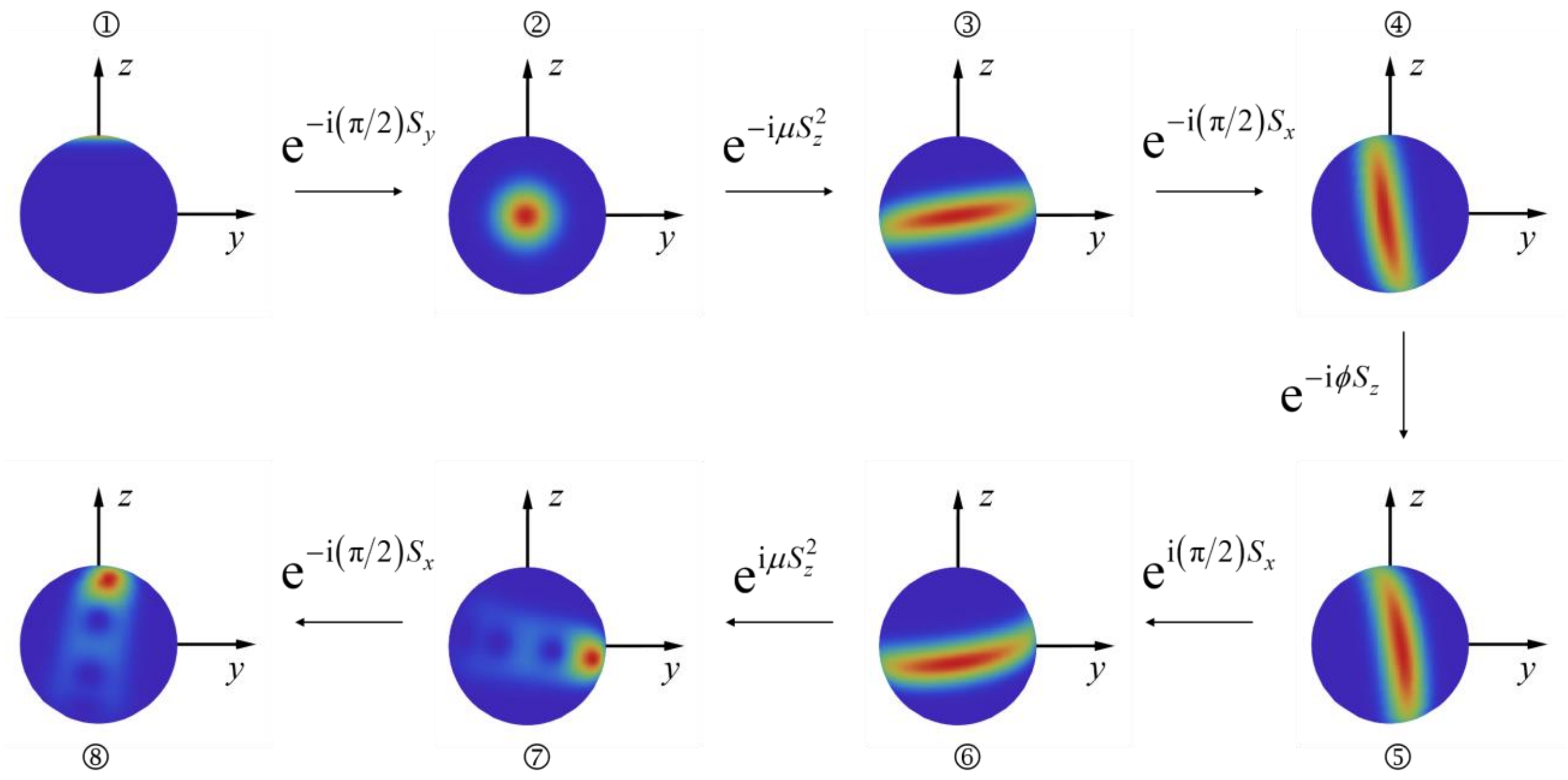

Figure 4. Complete conventional echo squeezing protocol (CESP) for the atomic clock. The process starts with the application of a π/2 pulse around the $y$-axis to the atoms in state $|\hat{\boldsymbol{z}}\rangle$, to produce the CSS $|\hat{\boldsymbol{x}}\rangle$. The effect of this step is equivalent to starting with the atoms in the CSS $|\hat{\boldsymbol{x}}\rangle$. The atoms then go through an OATS process with $\mu = \operatorname{arccot}\sqrt{N-2} \approx 1/\sqrt{N}$. This is followed in sequence by three steps: an auxiliary rotation around the $x$-axis, the rotation around the $z$-axis by an angle of $\phi$ due to the phase shift, and the reversal of the auxiliary rotation. These three steps are equivalent to a rotation around the $y$-axis by an angle of $\phi$. Next, the atoms go through an inverse OATS process. The net result of these additional four steps (namely, squeezing, auxiliary rotation, inversion of auxiliary rotation, and inverse squeezing) is a CSS state that would have been produced if the rotation around the $z$-axis were equal to a $\sim \phi\sqrt{N/\mathrm{e}}$ (for small $\phi$) without applying the additional four steps. As such, the net effect of the process is a phase magnification by a factor of $\sim \sqrt{N/\mathrm{e}}$. To measure this rotation around the $z$-axis, the CESP protocol uses a π/2 pulse around the $x$-axis, followed by a measurement of $S_z$. Measuring $S_z$ after the last π/2 pulse around the $x$-axis is equivalent to measuring $S_y$. It should be noted that the rotation axis of this last π/2 pulse is different from that of the first π/2 pulse, in contrast to the conventional Ramsey protocol, the SCSP, and the GESP.

It should be noted that there could be many equivalent variants of the simplified versions of the protocols because flipping the sign of the rotation angle of a π/2 pulse (for example, changing $R_y(-\pi/2)$ to $R_y(\pi/2)$) will only result in an additional minus sign in the final signal (this applies to all π/2 pulses, including the auxiliary pulse and its inverse). If we flip the signs of the rotation angles of two π/2 pulses, then the two minus signs in front of the final signal cancel out and the final signal will not be changed. Therefore, in practice, for all the π/2 pulses (including the auxiliary pulse and its inverse), we only need to specify the rotation axis, but do not need to care whether it is a clockwise or counterclockwise rotation.

The protocols have so far been illustrated only in the context of the Ramsey atomic clock above for concreteness. With proper modifications, these protocols can also be applied to the light pulse atom interferometer (LPAI). Here, we show in detail how to adapt these protocols to the LPAI [21,22,23]. This is important because the LPAI process involves additional steps, and it may not be a priori obvious how these steps may affect the overall behavior of any of the protocols. Recalling briefly, the conventional LPAI works as follows. The process starts with the pseudo-spins of all atoms pointing in the $+z$-direction. The wave packet of each atom is then split into two components by using a π/2 pulse of a pair of counter-propagating Raman beams. This step is formally equivalent to the application of a π/2 pulse in a Ramsey atomic clock. Each of the two components have different transverse momentum; as a result, the wave packet of the atoms gets spatially separated. After a dark period $T$, the two components interact again with the Raman beams, corresponding to a π pulse. After another dark period of $T$, another π/2 pulse is applied. This is followed by the detection of $S_z$. The phase shift in the LPAI can be due to various effect; for concreteness, we consider here the case where the phase shift is due to the Sagnac effect [24]. Furthermore, we assume that half of the phase accumulation occurs before the π pulse, while the remaining half occurs after the π pulse. Noting that $R_y(\pi) = -R_y(-\pi)$, the propagator describing the conventional LPAI protocol can be expressed as

$$
\begin{aligned}
&R_y\left(\frac{\pi}{2}\right)R_z\left(-\frac{\phi}{2}\right)R_y\left(-\pi\right)R_z\left(\frac{\phi}{2}\right)R_y\left(\frac{\pi}{2}\right)\\
&=\left[R_y\left(\frac{\pi}{2}\right)R_z\left(-\frac{\phi}{2}\right)R_y\left(-\frac{\pi}{2}\right)\right]\left[R_y\left(-\frac{\pi}{2}\right)R_z\left(\frac{\phi}{2}\right)R_y\left(\frac{\pi}{2}\right)\right]\\
&=R_x\left(-\frac{\phi}{2}\right)R_x\left(-\frac{\phi}{2}\right)=R_x\left(-\phi\right)
\end{aligned}
\tag{4}
$$

The first line in this equation expresses the propagator in a manner that corresponds to the actual physical steps. In going from the second line to the third line, we have made use of the rules stated in Eq. (2). Thus, we see that the whole LPAI operation can be viewed simply as a rotation around the *x*-axis by an angle of $-\phi$. Noting that the equivalent step for the conventional Ramsey atomic clock is also $R_y\left(-\pi/2\right)R_z\left(\phi\right)R_y\left(\pi/2\right)=R_x\left(-\phi\right)$ (here, we have made use of the fact that the sign of the rotation angle of any π/2 pulse does not affect the observable signal, as noted earlier), we can see that the single equivalent step of the conventional Ramsey clock is identical to that of the LPAI.

When the squeezing process is introduced, thus creating entangled states of the atoms, it may not be a priori clear how to account for the Sagnac effect in the LPAI. To address this issue, we reconsider first the case of the conventional LPAI in which the atoms are not entangled. In this case, we can simply consider each atom separately. The one-atom states $\left|\hat{\boldsymbol{z}}_0\right\rangle$ and $\left|-\hat{\boldsymbol{z}}_0\right\rangle$ both have well-defined trajectories, which will form a well-defined enclosed area. Therefore, it is obvious that the Sagnac phase shift gained by one atom is $\phi=2m\boldsymbol{\Omega}\cdot\boldsymbol{A}$ (still setting $\hbar=1$) where $m$ is the mass of one atom, $\boldsymbol{\Omega}$ is the rotational velocity of the LPAI, and $\boldsymbol{A}$ is the vectorial area enclosed by the two arms of the LPAI [24]. Accordingly, as we have done in Eq. (4), the Sagnac phase shift can be described by the propagator $\mathrm{e}^{-\mathrm{i}(\phi/2)S_z}$ in the first dark period and $\mathrm{e}^{\mathrm{i}(\phi/2)S_z}$ in the second dark period (the change in sign is due to the π pulse in the middle). For the case where atoms are entangled, we consider first the simplest case, namely the $z$-directed Schrödinger cat state generated, for example, in the SCSP-e protocol for even $N$ applied to the LPAI [24]. In this case, there are only two quantum states, namely $|\boldsymbol{z}\rangle$ and $|-\boldsymbol{z}\rangle$, each with a well-defined trajectory. Consequently, the Sagnac phase shift for the whole ensemble is simply $N\phi$, because the mass of each quantum state is $Nm$ [24]. This Sagnac phase shift still can be described by the propagator $\mathrm{e}^{-\mathrm{i}(\phi/2)S_z}$ in the first dark period and $\mathrm{e}^{\mathrm{i}(\phi/2)S_z}$ in the second dark period. This is affirmed by noting that the resulting effect agrees with the physical interpretation of the phase magnification stated above.

Consider next the case where the quantum state produced after the first squeezing pulse is not a simple Schrödinger cat state, but rather an arbitrary superposition of the symmetric Dicke states [25,26]. We recall that such a Dicke state is a superposition of many direct product states. A direct product state is an $N$-atom state where each atom is either in state $|\hat{\boldsymbol{z}}_0\rangle$ or $|-\hat{\boldsymbol{z}}_0\rangle$ but not a superposition. For such a direct product state, each atom has a well-defined trajectory. However, a direct product state as a whole does not have a well-defined trajectory because some atoms are in state $|\hat{\boldsymbol{z}}_0\rangle$ and follow one arm while the other atoms are in state $|-\hat{\boldsymbol{z}}_0\rangle$ and follow the other arm. In this case, we must consider the Sagnac phase shift contributed by each atom. In a Dicke state (except for the extremal ones), the atoms are entangled. Nonetheless, it is still true that for each atom, state $|\hat{\boldsymbol{z}}_0\rangle$ and $|-\hat{\boldsymbol{z}}_0\rangle$ will follow the two arms of the LPAI and gain different phases, whose difference is the overall Sagnac phase shift for the atom. This effect, for each atom, can still be described by the propagator $\mathrm{e}^{-\mathrm{i}(\phi/2)s_z}$ (noting that $s_z$ here is in lowercase, meaning the single atom operator) in the first dark period and $\mathrm{e}^{\mathrm{i}(\phi/2)s_z}$ in the second dark period. The Sagnac phase shift for the whole ensemble can then be described by the direct product of these single-atom propagators, namely $e^{-i(\phi/2)S_z}$ in the first dark period and $e^{i(\phi/2)S_z}$ in the second dark period; this follows from the definition of $S_z$ in Eq. (1). Not surprisingly, this formulation yields the correct result for the unentangled atoms as well for the case of the *z*-directed Schrödinger cat state.

The complete protocol of the GESP-e(o) for the LPAI can now be expressed as follows [24]: π/2 pulse around the *y*-axis applied to atoms in state $|\hat{\boldsymbol{z}}\rangle$, squeezing, auxiliary rotation around the *y*(*x*)-axis, the first dark period, π pulse around the *y*(*x*)-axis, the second dark period, the inverse auxiliary rotation, the inverse squeezing process, and a π/2 pulse around the *y*-axis. The final state expressed in terms of the propagator of the GESP-e(o) steps is shown in Eq. (5).

$$\left|\psi_{\substack{\mathrm{e}\\\mathrm{o}}}\right\rangle = R_y\left(\frac{\pi}{2}\right)\mathrm{e}^{\mathrm{i}\mu S_z^2}R_{\substack{y\\x}}\left(\mp\frac{\pi}{2}\right)R_z\left(-\frac{\phi}{2}\right)R_{\substack{y\\x}}(\pi)R_z\left(\frac{\phi}{2}\right)R_{\substack{y\\x}}\left(\mp\frac{\pi}{2}\right)\mathrm{e}^{-\mathrm{i}\mu S_z^2}R_y\left(\frac{\pi}{2}\right)|\hat{\boldsymbol{z}}\rangle \tag{5}$$

For $\mu = \pi/2$, Eq. (5) reproduces the result reported previously in Refs. [11] and [24] for the SCSP.

Ref. [1] does not explicitly consider the adaptation of the CESP protocol to the LPAI. However, following the same logic that was used for the CESP protocol for the Ramsey atomic clock, we can easily see that the CESP for the LPAI would be essentially identical to the GESP-o, with the only difference being that the last π/2 pulse is around the *x*-axis. Of course, just in the case of the Ramsey clock, GESP-e differs significantly from the CESP. It should be noted that the middle π pulse in the GESP-o and the CESP is around the *x*-axis, in contrast to the conventional LPAI and GESP-e.

The middle five steps for any of these protocols (the auxiliary rotation, the first dark period, the π pulse, the second dark period, and the inverse auxiliary rotation) is equivalent to a single rotation around the *x*- or *y*-axis, according to the relations

$$R_{\substack{x\\y}}(\phi)=R_{\substack{y\\x}}\left(\mp\frac{\pi}{2}\right)R_z\left(-\frac{\phi}{2}\right)R_{\substack{y\\x}}(\pi)R_z\left(\frac{\phi}{2}\right)R_{\substack{y\\x}}\left(\mp\frac{\pi}{2}\right) \tag{6}$$

It is now obvious that these complete protocols for the LPAI employing spin-squeezing are equivalent to the corresponding ones for the Ramsey atomic clock employing spin-squeezing shown earlier in Figure 3. As discussed above, for all the π/2 pulses (including the auxiliary pulse as well as its inverse), we only need to care about the rotation axis, but do not need to care about the sign of the rotation angle. Furthermore, obviously a π pulse is always equivalent to a $-\pi$-pulse around the same axis. Therefore, for any protocol, either for the Ramsey atomic clock or the LPAI, the sign of the rotation angle can be flipped for any of the pulses.

# 3. Sensitivity and the quantum Cramér-Rao bound of the GESP and the CESP

For both the Ramsey clock and the light pulse atom interferometer, the fundamental uncertainty in the measurement is directly proportional to the uncertainty in the measurement of the phase shift accumulated during the interaction. As such, the sensitivity of a protocol is characterized by the reciprocal of minimum measurable change in the phase, denoted as $\Delta\phi^{-1}$. The goal of any spin squeezing protocol is to maximize this parameter, which can be expressed in general as: $\Delta\phi^{-1}\big|_{\phi=0}=\left(\left|\partial_\phi\langle S_w\rangle\right|/\Delta S_w\right)_{\phi=0}$, where $S_w$ is the spin operator we measure, and $\Delta S_w$ is the standard deviation. Here, the numerator is referred to as the phase gradient and the denominator as the noise. The sensitivity of the CESP, considering only the quantum projection noise, has been analytically calculated [1] to be $\Delta\phi^{-1}\big|_{\phi=0}=\sqrt{2S}\,(2S-1)\cos^{2S-2}\mu\sin\mu$, with $S=N/2$, while the sensitivity of the GESP has only been investigated with numerical simulations [11,24]. The numerical simulations are constrained by three limitations. First, it is very difficult to simulate the result for values of $N$ larger than few hundreds. Second, little insight can be gained from the behavior determined for a limited range of values of $N$. Third, the actual mechanism underlying the enhancement of sensitivity, as an interplay between different degrees of phase magnification and amplification of the quantum projection noise, is not at all evident. To overcome these constraints, we have developed analytic expressions for determining the sensitivity of the GESP, as described next.

The final state of the three-step GESP shown in Figure 3 is

$$\left|\psi_{\substack{\mathrm{e}\\\mathrm{o}}}\right\rangle = \mathrm{e}^{\mathrm{i}\mu S_z^2}\mathrm{e}^{-\mathrm{i}\phi S_{\substack{x\\y}}}\mathrm{e}^{-\mathrm{i}\mu S_z^2}\left|\frac{\pi}{2},0\right\rangle = \mathrm{e}^{\mathrm{i}\mu S_z^2}\left(1-\mathrm{i}\phi S_{\substack{x\\y}}-\frac{1}{2}\phi^2 S_{\substack{x\\y}}^2+\mathcal{O}\left(\phi^3\right)\right)\mathrm{e}^{-\mathrm{i}\mu S_z^2}\left|\hat{\boldsymbol{x}}\right\rangle \tag{7}$$

The corresponding signal is then given by:

$$\begin{aligned}
\left\langle\psi_{\substack{\mathrm{e}\\\mathrm{o}}}\right|S_x\left|\psi_{\substack{\mathrm{e}\\\mathrm{o}}}\right\rangle &= \left\langle S_x\right\rangle - \mathrm{i}\phi\left\langle\left[S_x,\tilde{S}_{\substack{x\\y}}\right]\right\rangle + \phi^2\left\langle\left(\tilde{S}_{\substack{x\\y}}S_x\tilde{S}_{\substack{x\\y}}-\frac{1}{2}\left\{S_x,\tilde{S}_{\substack{x\\y}}^2\right\}\right)\right\rangle + \mathcal{O}\left(\phi^3\right)\\
&= S + \phi^2\left\langle\left(\tilde{S}_{\substack{x\\y}}S_x\tilde{S}_{\substack{x\\y}}-\frac{1}{2}\left\{S_x,\tilde{S}_{\substack{x\\y}}^2\right\}\right)\right\rangle + \mathcal{O}\left(\phi^4\right)\\
&= S + \phi^2\left(\left\langle\tilde{S}_{\substack{x\\y}}S_x\tilde{S}_{\substack{x\\y}}\right\rangle - S\left\langle\tilde{S}_{\substack{x\\y}}^2\right\rangle\right) + \mathcal{O}\left(\phi^4\right)
\end{aligned} \tag{8}$$

where, we define, for example, $\left\langle S_{\substack{x\\y}}\right\rangle \equiv \left\langle\hat{\boldsymbol{x}}\right|S_{\substack{x\\y}}\left|\hat{\boldsymbol{x}}\right\rangle$ and $\tilde{S}_{\substack{x\\y}} \equiv \mathrm{e}^{\mathrm{i}\mu S_z^2}S_{\substack{x\\y}}\mathrm{e}^{-\mathrm{i}\mu S_z^2}$. The two steps in Eq. (8) are both based on the fact that the CSS $\left|\hat{\boldsymbol{x}}\right\rangle$ is an eigenstate of $S_x$ with the eigenvalue of $S = N/2$. Here, the linear term proportional to $\phi$ vanishes as expected because the signal of the GESP is symmetric about $\phi = 0$. In fact, all odd order terms of $\phi$ vanish for the same reason. Therefore, in Eq. (8), the residual term in the second line becomes $\mathcal{O}\left(\phi^4\right)$. The details of the calculation of the signal are presented in Appendix B. The result is:

$$\left\langle \psi_{\substack{\mathrm{e}\\\mathrm{o}}} \middle| S_x \middle| \psi_{\substack{\mathrm{e}\\\mathrm{o}}} \right\rangle = S + \phi^2 \left( \boldsymbol{a}_{10} \cdot \boldsymbol{b}_0 \pm \boldsymbol{a}_{11} \cdot \boldsymbol{b}_1 \right) + \mathcal{O}\left(\phi^4\right) \tag{9}$$

where the expressions of the vectors $\boldsymbol{a}_{10}$, $\boldsymbol{a}_{11}$, $\boldsymbol{b}_0$, and $\boldsymbol{b}_1$ are given in Appendix B. Applying the same approach as outlined above for deriving the signal, we find the standard deviation to be:

$$\Delta S_x^2 \equiv \left\langle \psi_{\substack{\mathrm{e}\\\mathrm{o}}} \middle| S_x^2 \middle| \psi_{\substack{\mathrm{e}\\\mathrm{o}}} \right\rangle - \left\langle \psi_{\substack{\mathrm{e}\\\mathrm{o}}} \middle| S_x \middle| \psi_{\substack{\mathrm{e}\\\mathrm{o}}} \right\rangle^2 = \phi^2 S^4 \left( \boldsymbol{a}_{30} \cdot \boldsymbol{b}_0 \pm \boldsymbol{a}_{31} \cdot \boldsymbol{b}_1 \right) + \mathcal{O}\left(\phi^4\right) \tag{10}$$

The detail of steps in the derivation of this result as well as the expressions for the vectors $\boldsymbol{a}_{30}$ and $\boldsymbol{a}_{31}$ are also shown in Appendix B.

If we substitute Eqs. (9) and (10) into $\Delta\phi^{-1}\Big|_{\phi=0} = \left( \left| \partial_\phi \left\langle S_{\substack{x\\y}} \right\rangle \right| \Big/ \Delta S_{\substack{x\\y}} \right)_{\phi=0}$, we find that both the numerator and the denominator are zero. Therefore, we calculate the sensitivity by applying the L'Hôpital's rule:

$$\Delta\phi_{\substack{\mathrm{e}\\\mathrm{o}}}^{-1}\Big|_{\phi=0} = \lim_{\phi\to 0} \frac{\left|\partial_\phi \left\langle S_x \right\rangle\right|}{\Delta S_x} = \frac{\partial_\phi \left|\partial_\phi \left\langle S_x \right\rangle\right|}{\partial_\phi \Delta S_x}\Bigg|_{\phi=0} = \frac{2\left|\boldsymbol{a}_{10} \cdot \boldsymbol{b}_0 \pm \boldsymbol{a}_{11} \cdot \boldsymbol{b}_1\right|}{\sqrt{\boldsymbol{a}_{30} \cdot \boldsymbol{b}_0 \pm \boldsymbol{a}_{31} \cdot \boldsymbol{b}_1}} \tag{11}$$

It should be noted that because the phase gradient and the quantum projection noise of the GESP both vanish at $\phi = 0$, the actual sensitivity in the presence of detection noise will also vanish for this phase shift. To solve this problem, we should operate the GESP at a non-zero phase shift, which is discussed in detail in Sec. 5.

The quantum Cramer-Rao (QCR) bound determines the maximum sensitivity that can be achieved, thereby representing the optimal retrieval of the quantum Fisher information present in the system [27]. For a pure state, the QCR bound for phase sensitivity is given by $\Delta\phi^{-1}\big|_{\phi=0} \le 2\Delta S_{\hat{\boldsymbol{n}}}$, where $\hat{\boldsymbol{n}}$ is the axis of the rotation representing the phase shift, $\boldsymbol{S}_{\hat{\boldsymbol{n}}} \equiv \boldsymbol{S}\cdot\hat{\boldsymbol{n}}$, and $\Delta S_{\hat{\boldsymbol{n}}}$ is the standard deviation of $S_{\hat{\boldsymbol{n}}}$ for the state just before introducing the phase shift. Using the result of the standard deviation after one-axis-twist squeezing process shown in Ref. [3], we can write the QCR bound for the GESP-e as

$$\Delta\phi^{-1}\big|_{\phi=0} \le 2\Delta S_x = \sqrt{2S\left(S+\frac{1}{2}\right)+2S\left(S-\frac{1}{2}\right)\cos^{2S-2}2\mu-\left(2S\cos^{2S-1}\mu\right)^2} \tag{12}$$

and the QCR bound for the GESP-o and CESP as

$$\Delta\phi^{-1}\big|_{\phi=0} \le 2\Delta S_y = \sqrt{2S\left(S+\frac{1}{2}\right)-2S\left(S-\frac{1}{2}\right)\cos^{2S-2}2\mu} \tag{13}$$

The sensitivity and the QRC bound of the GESP-e, GESP-o, and the CESP for $N$ = 100 and 101 are plotted in Figure 5. In each case, it can be seen that the sensitivity of the GESP is close to the QRC bound over the whole range of $\mu$, while the sensitivity of the CESP is close to the QRC bound in a small interval of $\mu$ near zero. These behaviors of the sensitivities mean that the CESP wastes almost all quantum Fisher information except for small values of $\mu$, while both versions of the GESP makes good use of the quantum Fisher information. It should be noted that the substantial difference in behavior between the CESP and GESP-o results only from the quantum operator measured in the end. To give an insight into why this minor difference in protocol can result in a substantial difference in behavior, we plot the Husimi quasi-probability distribution of the quantum state in each step of the CESP, which is also the quantum state in each step of GESP-o, for $\mu \gg 1/\sqrt{N}$ , in Figure 6. In this regime, the final state is always symmetric about the *x*-*z* plane for small phase shifts. Therefore, $\langle S_y \rangle \approx 0$ and does not indicate the phase shift. In contrast, the phase shift will still change $\langle S_x \rangle$ significantly. Consequently, by measuring $\langle S_x \rangle$, we still can obtain the phase shift. Other advantages of the GESP will be discussed in Section 4.

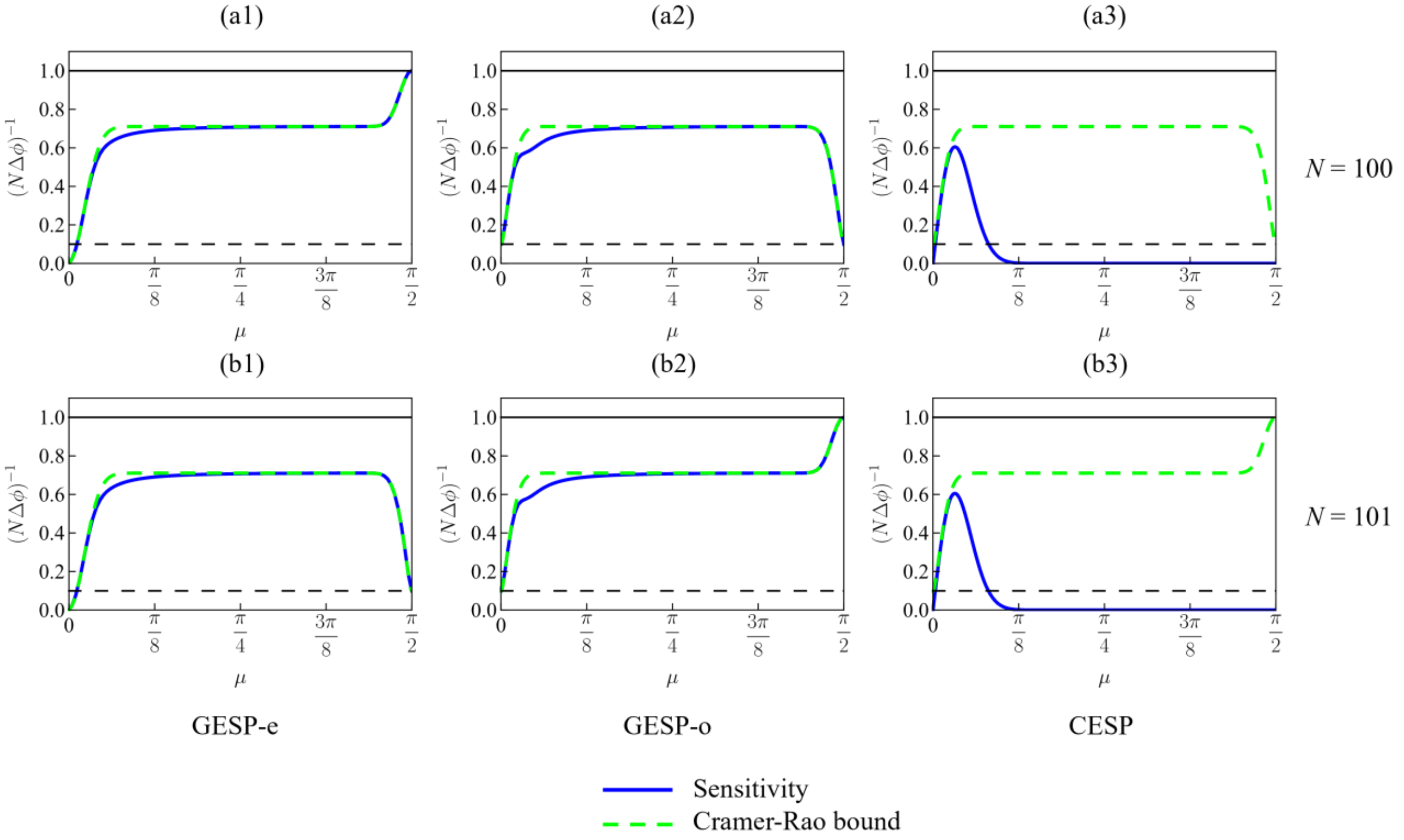

Figure 5. Sensitivity (blue solid) and the quantum Cramér-Rao bound (green dashed), normalized to the Heisenberg limit, for three different protocols, for $N = 100$ (top row) and $N = 101$ (bottom row). The first column is for GESP-e, the second column is for GESP-o, and the third column is for the CESP. In each plot, the solid black line is the Heisenberg limit, and the dashed black line is the standard quantum limit. It can be seen that the sensitivity of the GESP is close to the QRC bound over the whole range of $\mu$ , while the sensitivity of the CESP is close to the QRC bound in a small interval of $\mu$ near zero. These behaviors of the sensitivities mean that the CESP wastes almost all quantum Fisher information except for small values of $\mu$ , while both versions of the GESP makes good use of the quantum Fisher information.

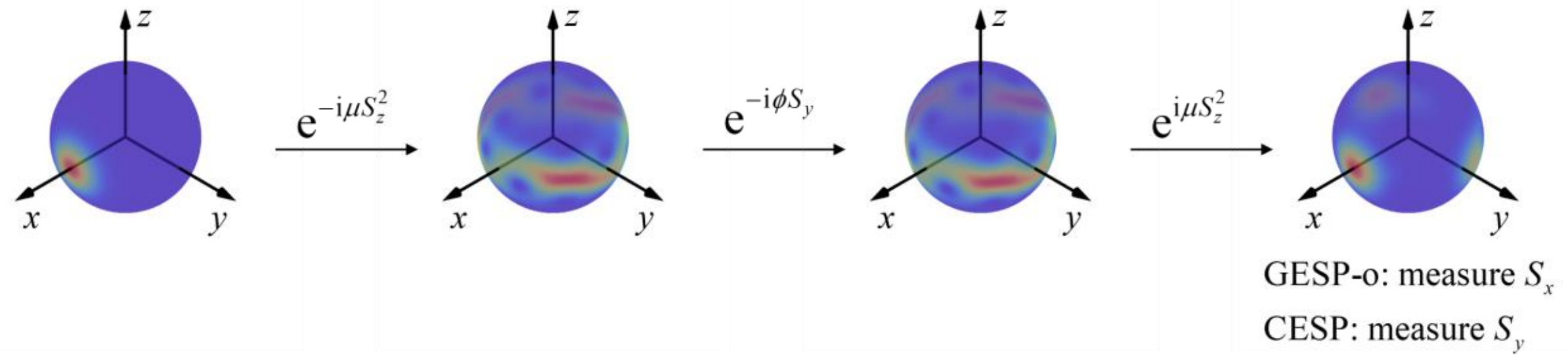

Figure 6 Quantum state in each step of the CESP, which is also the quantum state in each step of GESP-o. In the regime $\mu \gg N^{-1/2}$, the final state is always symmetric about the *x*-*z* plane for small phase shifts. Therefore, $\langle S_y \rangle \approx 0$ and does not indicate the phase shift. In contrast, the phase shift will still change $\langle S_x \rangle$ significantly. Consequently, by measuring $\langle S_x \rangle$, we still can obtain the phase shift.

# 4. Properties of the sensitivity of the GESP

It can be seen from Figure 5 that the sensitivity of the GESP, as expressed in Eq. (11), first increases as $\mu$ increases from 0, and remains almost a constant until $\mu$ approaches π/2, and then either increases to the Heisenberg limit or decreases to the standard quantum limit depending on the pairing of the parity of $N$ and the protocol as discussed in Section 2. The middle interval where the sensitivity is almost constant with respect to $\mu$ is referred to as the plateau region, which is the interval the GESP should be operated in. In this plateau, the sensitivity does not show the dependence on the parity of $N$.

We next determine quantitatively the boundaries of this plateau. We first notice that the right boundary of the plateau is the point where the sensitivity starts to depend on the parity of $N$. The dependence on the parity of $N$ for values of $\mu$ close to $\pi/2$, turns out to a consequence of the behavior of the quantity $\cos^{2S} 2\mu$ in vector $\boldsymbol{b}_1$ in Eq. (11). This quantity starts to demonstrate a strong parity dependence as $\mu$ approaches $\pi/2$, as shown in Figure 7. Therefore, the right boundary of the plateau can be determined by solving the equation $\cos^{2S} \approx 0$. Here, for computational convenience, we assume that $\cos^{2S} \approx 0$ if its absolute value is less than $\mathrm{e}^{-4} \approx 0.018$. Solving the inequality $\cos^{2S} 2\mu \le \mathrm{e}^{-4}$, we find this range to be $S^{-1/2} \le \mu \le \left(\pi/2 - S^{-1/2}\right)$ for $S \gg 1$. Therefore, the right boundary of the plateau is $\mu = \left(\pi/2 - S^{-1/2}\right)$.

To find the left boundary of the plateau, , we substitute $\cos^{2S} 2\mu$ with 0 in Eq. (11) for the interval $S^{-1/2} \le \mu \le \left(\pi/2 - S^{-1/2}\right)$, which yields:

$$\Delta\phi_{\substack{\mathrm{e}\\\mathrm{o}}}^{-1}\Big|_{\phi=0} \approx \frac{2\left(\boldsymbol{a}_{10} \cdot \boldsymbol{b}_0\right)}{\sqrt{\boldsymbol{a}_{30} \cdot \boldsymbol{b}_0}}$$
$$= \frac{\sqrt{2}S\left[\left(1-\cos 2\mu\right) + S^{-1}\frac{1}{2}\left(1-\cos 2\mu\right) + S^{-2}\frac{1}{2}\cos 2\mu\right]}{\sqrt{\left(1-\cos 2\mu\right)^2 + S^{-1}\left(1-\cos 2\mu\right) + S^{-2}\left(\frac{3}{4} + \cos 2\mu - \frac{7}{4}\cos^2 2\mu\right) - S^{-3}\frac{1}{4}\left(1-3\cos^2 2\mu\right)}} \tag{14}$$

If the condition that $\left(1-\cos 2\mu\right)^2 \gg S^{-1}\left(1-\cos 2\mu\right)$ is satisfied, this expression can be further simplified to give:

$$\Delta\phi_{\substack{\mathrm{e}\\\mathrm{o}}}^{-1}\Big|_{\phi=0} \approx \frac{\sqrt{2}S\left(1-\cos 2\mu\right)}{\sqrt{\left(1-\cos 2\mu\right)^2}} = \sqrt{2}S = \frac{N}{\sqrt{2}} \tag{15}$$

The result of Eq. (15) shows that in the interval defined by the conditions $S^{-1/2} \le \mu \le \left(\pi/2 - S^{-1/2}\right)$ and $\left(1-\cos 2\mu\right)^2 \gg S^{-1}\left(1-\cos 2\mu\right)$, the sensitivity of the GESP is the Heisenberg limit within a factor of $\sqrt{2}$, which does not depend on $\mu$, and thus corresponds to the plateau. Therefore, the left boundary of the plateau can be found by solving the inequality $\left(1-\cos 2\mu\right)^2 \gg S^{-1}\left(1-\cos 2\mu\right)$. We again make a computationally convenient specification that this inequality is satisfied if $\left(1-\cos 2\mu\right)^2 \ge 32S^{-1}\left(1-\cos 2\mu\right)$, which implies that $\left(1-\cos 2\mu\right) \ge 32S^{-1}$. It is easy to see from this inequality that for a very large value of $S$, the value of $2\mu$ has to be very small, so that we can expand $\cos 2\mu$ to second order in $\mu$ around $\mu = 0$. We then get the simple expression that $\mu \geq 4S^{-1/2}$. Combining this with the range of validity established in producing Eq. (14), we thus get the result that for $4S^{-1/2} \leq \mu \leq (\pi/2 - S^{-1/2})$, the sensitivity of the GESP is approximately $N/\sqrt{2}$. This approximated plateau region agrees closely with exact evaluations of the sensitivity using Eq. (11), as shown in Figure 8(a1) and (b1). This property of the sensitivity of the GESP means that for large *N*, we can work at almost any value of $\mu$ except for narrow intervals near $\mu = 0$ and $\mu = \pi/2$, to obtain a sensitivity that is within a factor of $\sqrt{2}$ of the Heisenberg limit, as also shown in Figure 8(a1) and (b1). On the contrary, the peak of the sensitivity curve of the CESP becomes very narrow for a large *N*. The sensitivity of the CESP becomes $\sim 0$ when $\cos^{2S}\mu \approx 0$ according to its analytical expression shown earlier. This means that the sensitivity of the CESP is greater than 0 only in the interval $0 < \mu < 2S^{-1/2}$. Therefore, when using the CESP, one must tune the value of $\mu$ very precisely in an experiment to obtain an enhancement of the sensitivity.

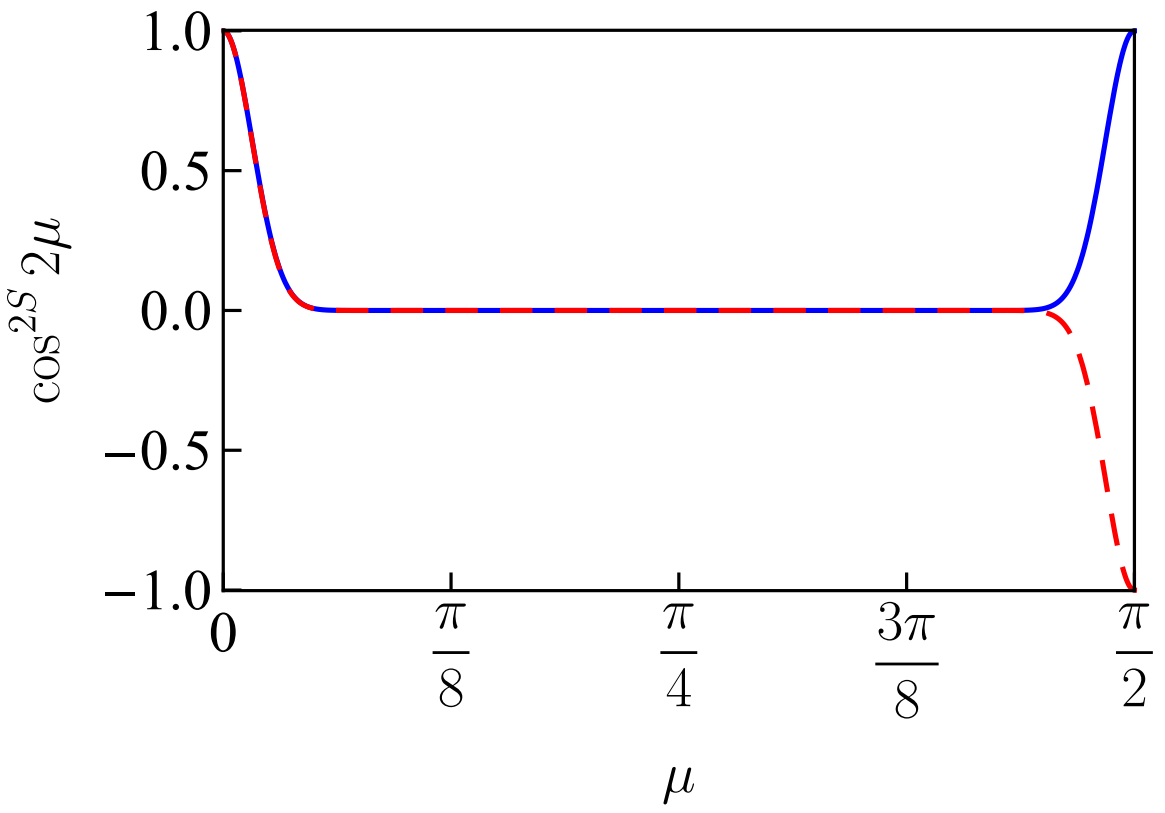

Figure 7. Plot of $\cos^{2S} 2\mu$ for $N = 100$ (solid blue) and $N = 101$ (dashed red). Its value decreases from 1 to $\sim 0$ as $\mu$ increases from 0 to $\sim 1/\sqrt{S}$, then remains $\sim 0$ until $\mu = (\pi/2 - 1/\sqrt{S})$, and finally increase to 1 or decrease to $-1$ depending on the parity of $N$.

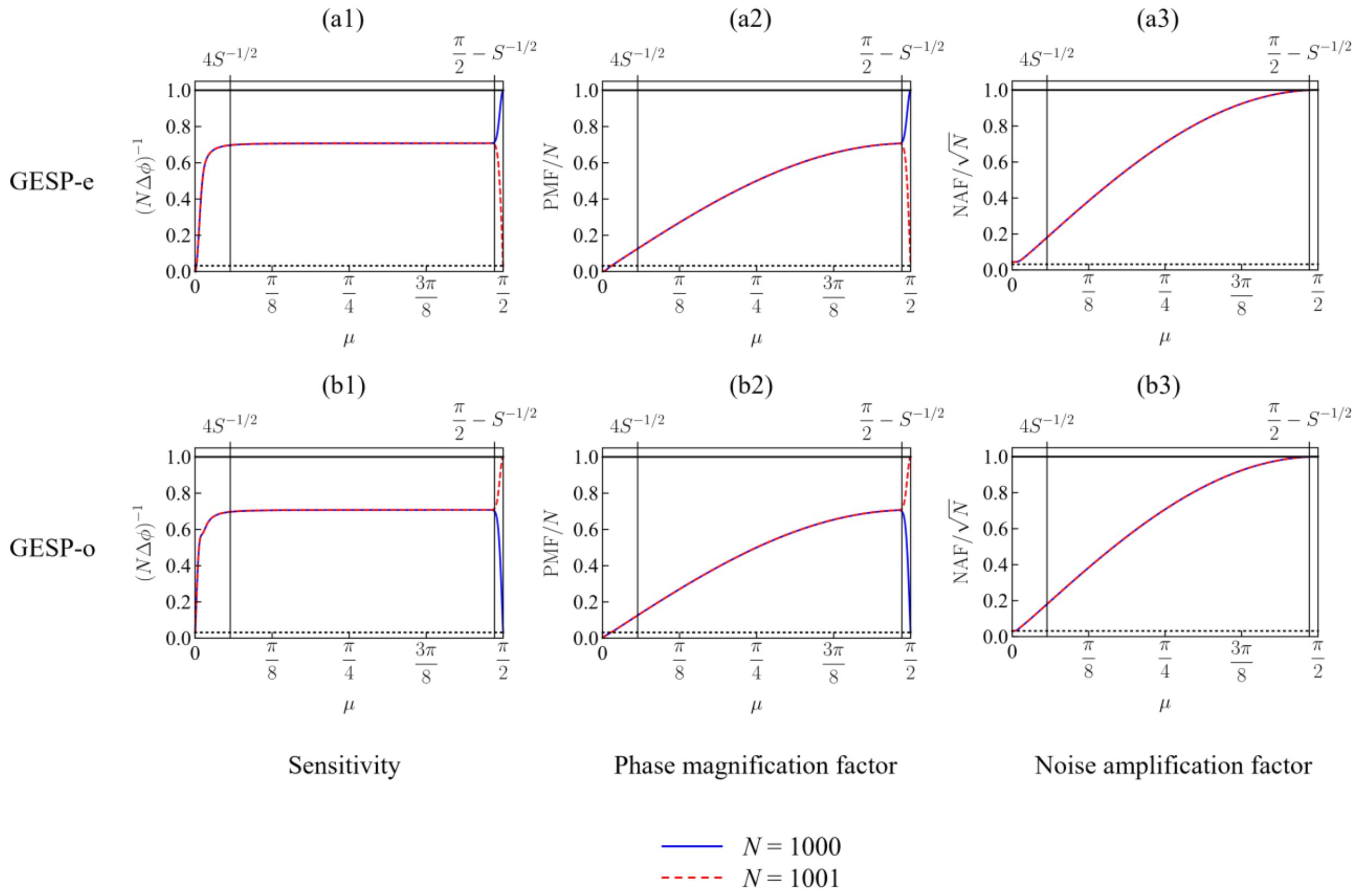

Figure 8. Sensitivity (first colume), phase magnification factor (PMF) (second colume), and noise amplification factor (NAF) (third colume) of GESP-e (first row) and GESP-o (second row) for $N = 1000$ (blue solid) and 1001 (red dashed). The sensitivity is normalized to the Heisenberg limit. The PMF is normalized to $N$. The NAF is normalized to $\sqrt{N}$. The two vertical lines in each plot mark the interval $4S^{-1/2} \le \mu \le \left(\pi/2 - S^{-1/2}\right)$ where the sensitivity is approximately the Heisenbergy limit divided by $\sqrt{2}$. In this interval, the PMF is approximately $\sqrt{2}S\sin\mu$ and the NAF is approximately $\sqrt{2S}\sin\mu$. In each plot, the horizontal solid black line is at the level of unity and the dashed line at the level of $1/\sqrt{N}$. The enhancement of the robustness against detection noise is given by the NAF.

# 5. Phase magnification and noise amplification

While the sensitivity analyzed in Section 3 indicates the limit of what can be achieved, it is also useful to determine the interplay between phase magnification and noise amplification in the determination of the net enhancement in sensitivity because the these two factors will determine the enhancement of sensitivity in the presence of detection noise [11]. To see why, note that the sensitivity including detection noise can be expressed as:

$$\Delta\phi^{-1}\Big|_{\phi=0} = \frac{\left|\partial_\phi \left\langle S_{x(y)} \right\rangle\right|}{\sqrt{\Delta S_{\mathrm{QPN}}^2 + \Delta S_{\mathrm{DN}}^2}}\Bigg|_{\phi=0} \tag{16}$$

where QPN stands for "quantum projection noise" and DN stands for "detection noise". In conventional use of OATS, the quantum projection noise is suppressed while the phase shift is not magnified. As such, such a process will not enhance the sensitivity significantly if detection noise is comparable or higher than the quantum projection noise of an unsqeezed state also with $N$ atoms. In the case of the CESP, the phase shift if magnified by a factor of $\sqrt{N/\mathrm{e}}$ without any variation in the quantum projection noise. Therefore, the sensitivity will be enhanced by a factor of $\sqrt{N/\mathrm{e}}$ but will still be far below the Heisenberg limit if detection noise is much higher than the quantum projection noise of an unsqueezed ensemble also with $N$ atoms. In the case of the SCSP, both the phase magnification and the quantum projection noise amplification are maximized, thus making it very robust against detection noise. As such, it is imperative to understand the degree of phase magnification and the quantum projection noise amplification for the GESP.

For a protocol that does not make use of spin squeezing, the signal is $S\cos\phi$, and the noise is $\sqrt{S/2}\,|\sin\phi|$. For a general case, in the proximity of $\phi=0$, any signal symmetric about $\phi=0$ can be approximated by $S\cos(M\phi)$ and the noise by $A\sqrt{S/2}\,|\sin(M\phi)|$, where the parameter $M$ represents the phase magnification factor (PMF), and the parameter $A$ represents the noise amplification factor (NAF). The net enhancement of the sensitivity is given by $(M/A)$ [11]. In the case of a protocol without spin squeezing, we have $M=A=1$. For the case where the signal is anti-symmetric about $\phi=0$, it is also possible to define equivalent values of the PMF and the

NAF, as we will show later. We now use this generalized framework to compare the spin-squeezing protocols described in this paper.

Consider first the PMF and the NAF for the two versions of the SCSP. For clarity of discussion, we introduce here the following terminology. The cases when SCSP-e is applied for even $N$ or SCSP-o is applied to odd $N$ are denoted as parity-matched SCSPs. On the other hand, the cases when SCSP-e is applied to odd $N$ or SCSP-o is applied to even $N$ are denoted as parity-crossed SCSPs. For the parity-matched SCSPs, the signal is $S\cos 2S\phi$, and the noise is $S\left|\sin 2S\phi\right|$ [28]. Thus, for the parity-matched SCSPs, the PMF is $M = 2S = N$ and the NAF is $A = \sqrt{2S} = \sqrt{N}$. The net enhancement of the sensitivity is then $(M/A) = \sqrt{N}$, reaching the Heisenberg limit. For parity-crossed SCSPs, the signal and the noise are a little more complicated. For the SCSP-o applied to even $N$, the final state is [17]

$$\left|\psi\right\rangle = \mathrm{e}^{\mathrm{i}\mu S_z^2}\mathrm{e}^{-\mathrm{i}\phi S_y}\mathrm{e}^{-\mathrm{i}\mu S_z^2}\left|\frac{\pi}{2},0\right\rangle = \frac{1}{2}\left[\left(\left|\frac{\pi}{2}-\phi,0\right\rangle + \left|\frac{\pi}{2}+\phi,0\right\rangle\right) + (-1)^S\,\mathrm{i}\left(\left|\frac{\pi}{2}+\phi,\pi\right\rangle - \left|\frac{\pi}{2}-\phi,\pi\right\rangle\right)\right] \quad (17)$$

The signal is thus given by:

$$\begin{aligned}\langle\psi|S_x|\psi\rangle &= \langle\psi|R_y\left(\frac{\pi}{2}\right)S_z R_y\left(-\frac{\pi}{2}\right)|\psi\rangle \\ &= \frac{1}{4}\begin{bmatrix}\left(\langle-\phi,0| + \langle\phi,0|\right) \\ -(-1)^S\,\mathrm{i}\left(\langle\pi+\phi,\pi| - \langle\pi-\phi,\pi|\right)\end{bmatrix} S_z \begin{bmatrix}\left(|-\phi,0\rangle + |\phi,0\rangle\right) \\ +(-1)^S\,\mathrm{i}\left(|\pi+\phi,\pi\rangle - |\pi-\phi,\pi\rangle\right)\end{bmatrix} \\ &= \frac{1}{2}\left(\mathrm{Re}\left(\langle-\phi,0|S_z|\phi,0\rangle\right) - \mathrm{Re}\left(\langle\pi+\phi,\pi|S_z|\pi-\phi,\pi\rangle\right)\right) \\ &= S\cos^{2S-1}\phi\end{aligned} \quad (18)$$

Using a similar analysis, it can be shown that the result of Eq. (18) is also valid for the SCSP-e applied to odd $N$. The interference fringes of the SCSP-e and the SCSP-o for $N$ = 20 and 21 are shown in Figure 9. In the proximity of $\phi = 0$, the signal for the parity-crossed SCSPs can be expressed as

$$\langle\psi|S_x|\psi\rangle = S\left[1+\frac{1}{2}\left(\phi\sqrt{2S-1}\right)^2\right]+\mathcal{O}\left(\phi^4\right) = S\cos\left(\phi\sqrt{2S-1}\right)+\mathcal{O}\left(\phi^4\right) \tag{19}$$

which agrees with the result shown in Eq. (9). The noise in the proximity of $\phi = 0$ calculated with Eq. (10) is

$$\Delta S_x = \sqrt{\frac{S}{2}}(2S-1)|\phi|+\mathcal{O}\left(\phi^3\right) = \sqrt{(2S-1)\frac{S}{2}}\left|\sin\left(\phi\sqrt{2S-1}\right)\right|+\mathcal{O}\left(\phi^3\right) \tag{20}$$

Therefore, both the PMF and the NAF are $\sqrt{2S-1}$, and it is approximately $\sqrt{N}$ in the limit $N \gg 1$. Consequently, there is no enhancement of the sensitivity for the parity-crossed SCSP. It should also be noted that the NAFs for both the parity-matched and parity-crossed SCSPs are approximately the same.

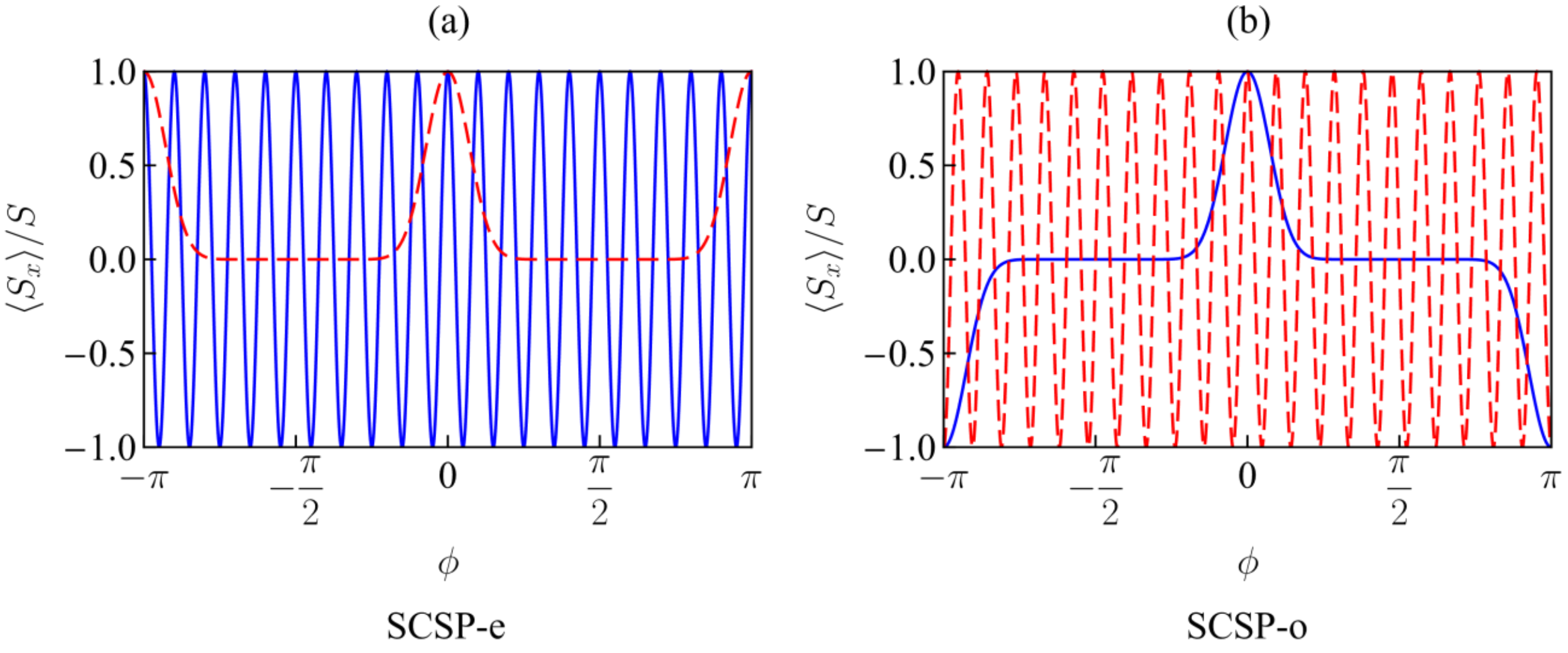

Figure 9. Signal of (a) the SCSP-e and (b) the SCSP-o for $N$ = 20 (blue solid) and $N$ = 21 (red dashed). The signal is $S\cos(2S\phi)$ for the parity-matched cases, and $S\cos^{2S-1}\phi$ for the parity-crossed cases.

Consider next the case of the CESP, for the optimal value of $\mu$, which is [1] $\operatorname{arccot}\sqrt{2S-2}$. In the proximity of $\phi=0$, the signal of the CESP is $\langle S_y\rangle = \left[\phi S\sin\mu\cos^{2S-2}\mu + \mathcal{O}\left(\phi^3\right)\right]$. In the limit $S \gg 1$, the optimal value of $\mu$ is approximately $1/\sqrt{2S} = 1/\sqrt{N}$ and the signal is:

$$\langle S_y\rangle = \phi S\sqrt{\frac{2S}{\mathrm{e}}} + \mathcal{O}\left(\phi^3\right) = S\sin\left(\phi\sqrt{\frac{2S}{\mathrm{e}}}\right) + \mathcal{O}\left(\phi^3\right) \tag{21}$$

Although the signal of the CESP is anti-symmetric, it is obvious from Eq. (21) that the effective PMF is $M = \sqrt{N/\mathrm{e}}$. The noise in the proximity of $\phi=0$ is:

$$\Delta S_y = \sqrt{\frac{S}{2}} + \mathcal{O}\left(\phi^2\right) = \sqrt{\frac{S}{2}}\cos M\phi + \mathcal{O}\left(\phi^2\right) \tag{22}$$

Thus, the noise is not amplified at all. Of course, it is evident that these results for the CESP does not depend on the parity of *N*. For the PMF, this is also evident in Figure 10 where we have shown the actual signals for the CESP for two different parities of *N*, without any approximations.

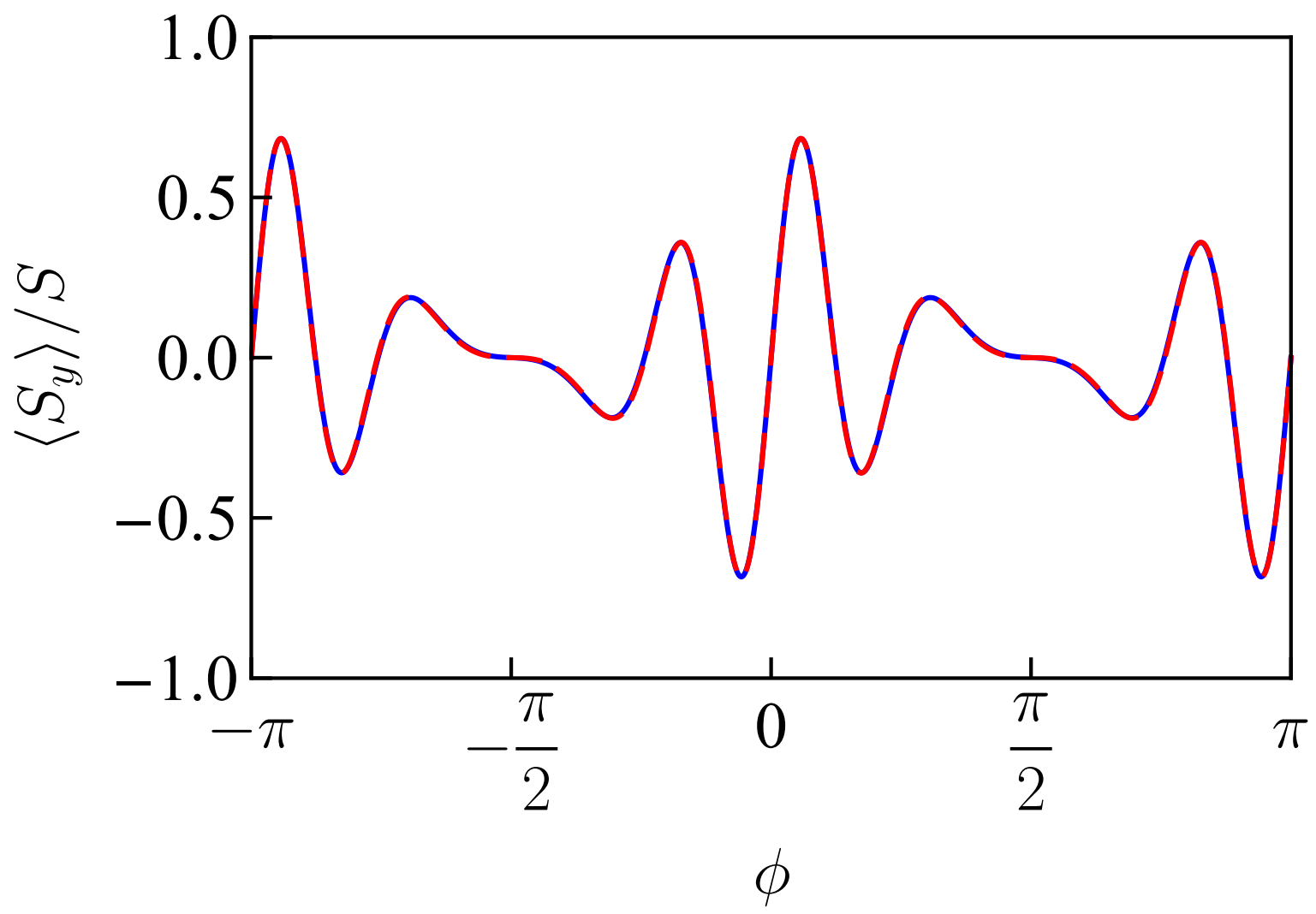

Figure 10. Signal of the CESP for $N$ = 100 (blue solid) and $N$ = 101 (red dashed) at the optimal value of the squeezing parameter $\mu = \operatorname{arccot}\sqrt{2S-2}$. The signal does not depend on the parity of $N$ for any value of $\phi$. The PMF in the proximity of $\phi = 0$ is $\sim\sqrt{N/\mathrm{e}}$.

Consider next the role of phase magnification and noise amplification in the case of the GESP. According to the discussion in Section 4, in the range of the plateau $4S^{-1/2} \le \mu \le \left(\pi/2 - S^{-1/2}\right)$ and for $S \gg 1$, the signal and the noise for the GESP described by Eq. (9) and Eq. (10) can be expressed as:

$$\left\langle \psi_{\mathrm{e(o)}} \middle| S_x \middle| \psi_{\mathrm{e(o)}} \right\rangle = S\left[1 - \frac{1}{2}\left(\sqrt{2}\phi S \sin\mu\right)^2\right] + \mathcal{O}\left(\phi^3\right) = S\cos\left(\sqrt{2}\phi S \sin\mu\right) + \mathcal{O}\left(\phi^4\right) \tag{23}$$

$$\Delta S_x = \sqrt{2}S^2\left|\phi\right|\sin^2\mu + \mathcal{O}\left(\phi^3\right) = \sqrt{\frac{S}{2}2S}\,\sin\mu\left|\sin\left(\sqrt{2}\phi S\sin\mu\right)\right| + \mathcal{O}\left(\phi^3\right) \tag{24}$$

Compared to the signal and the noise of the protocol without spin squeezing, we can see that for the GESP, the PMF is $M = \sqrt{2}S\sin\mu$ and the NAF is $A = \sqrt{2S}\sin\mu$. Therefore, the net enhancement of sensitivity is $\left(M/A\right) = \sqrt{S} = \sqrt{N/2}$, yielding the Heisenberg limit within a factor of $\sqrt{2}$. It should be noted that Eqs. (23) and (24) are valid for both even $N$ and odd $N$ cases. The signal of the GESP for $N = 100$ (blue solid) and 101(red dashed) are shown in Figure 11, without any approximations. As can be seen, the central fringe of the signal does not depend on the parity of *N*.

Taking into account the effect of detection noise, the sensitivity of the GESP in the range of the plateau can be expressed as

$$\Delta\phi^{-1}\Big|_{\phi=0} = \frac{\sqrt{2}S^2 \sin\mu \sin M\phi}{\sqrt{\left(S\sin\mu\sin M\phi\right)^2 + \Delta S_{\mathrm{DN}}^2}}\Bigg|_{\phi=0} \tag{25}$$

However, this expression has the problem that the numerator is zero at $\phi = 0$, so that the sensitivity vanishes in the presence of any detection noise. In fact, this problem emerges with any sensing mechanism that produces a signal that is symmetric around $\phi = 0$. To circumvent this problem for such a system, it is customary to employ the so-called hopping technique [29]. Under this technique, one uses the difference in the signals measured at two non-zero values of $\phi$ symmetric about $\phi = 0$. Typically, the measurements are made for values of $\phi$ where $\left|\partial_\phi \left\langle S_x \right\rangle\right|$ is maximum. According to Eq. (23), $\partial_\phi \left\langle S_x \right\rangle$ becomes the maximum at $\phi = \pm\pi/2M$. Since the analytic results are approximate, it is not a priori obvious whether the assumption used in deriving the results of Eqs. (23) and (24) are valid for these values of $\phi$. To check this, we have compared the results of Eqs. (23) and (24) with the exact simulation results. These are shown in Figure 12 for $N = 100$ and $\mu = \pi/4$. As can been seen, the results of Eqs. (23) and (24) represent good approximations of the actual signal and noise for $\phi = \pm\pi/2M$, and only deviates for significantly larger values of $\phi$. Additional simulations we have carried out (not shown) indicates that the agreement between the results of Eqs. (23) and (24) and the numerically calculated signal and noise increases with increasing value of $\mu$ .

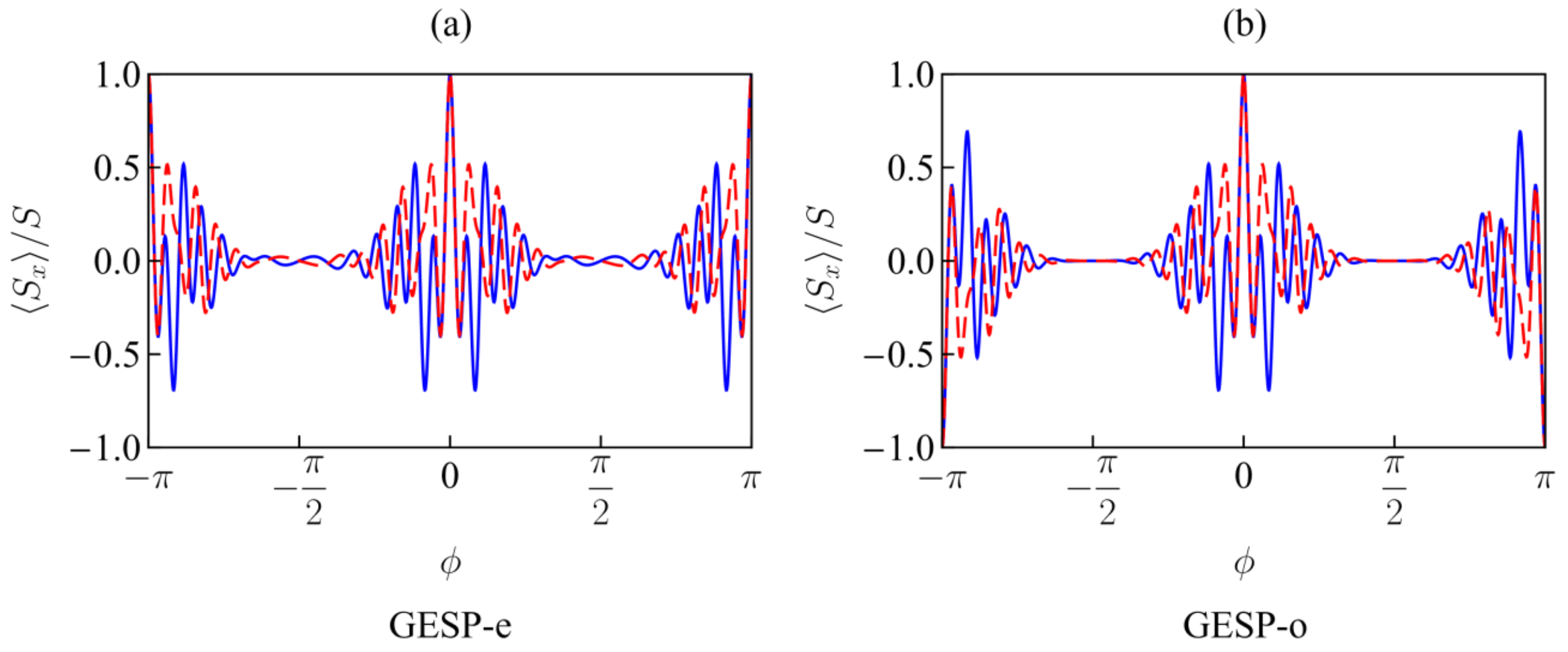

Figure 11. Signal of the (a) GESP-e and (b) GESP-o for $N$ = 100 (blue solid) and $N$ = 101 (red dashed) at $\mu = \pi/8$ . The central fringe of the signal does not depend on the parity of $N$. The fringes are narrower than that of the CESP, indicating a larger PMF for the GESP.

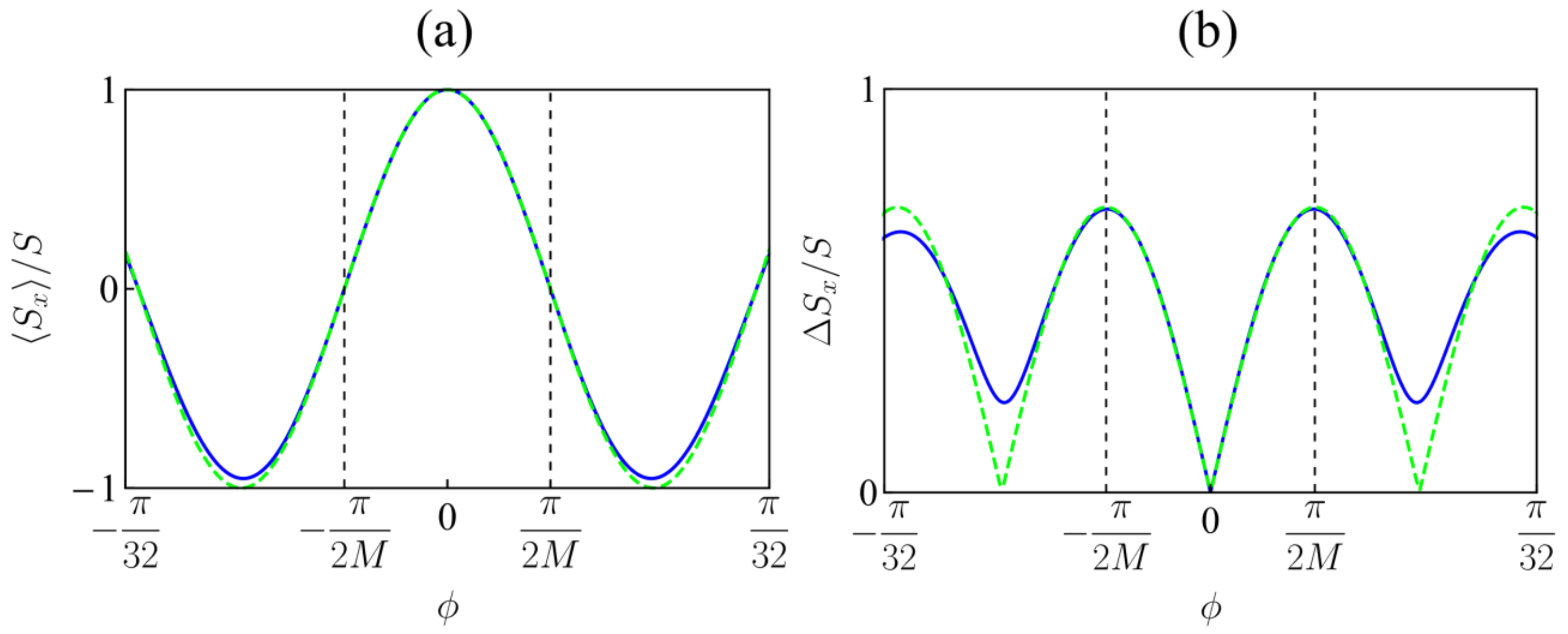

Figure 12. Comparison between the results in Eqs. (23) and (24) (blue solid) with the simulation results without approximations (green dashed). (a) Signal (b) Quantum projection noise (robustness). For the central fringe, the Eqs. (23) and (24) remain good approximations of the simulation results.

The signal observed under the hopping technique, with a swing parameter of $\pm\pi/2M$, can be expressed as the expectation value of $S_{\text{hop}}(\varepsilon) = \left[ S_x|_{\phi=\varepsilon+\pi/2M} - S_x|_{\phi=\varepsilon-\pi/2M} \right] / 2$, where $\varepsilon$ represents the phase shift being measured. Considering that the signal for the GESP is symmetric around $\phi = 0$, and the noises for different values of $\phi$ are uncorrelated, we can express the phase gradient of the signal as $\partial_\varepsilon \langle S_{\text{hop}} \rangle = \partial_\phi \langle S_x \rangle|_{\phi=\pi/2M}$ and the noise as $\Delta S_{\text{hop}} = \Delta S_x|_{\phi=\pi/2M}$. Therefore, using the hopping technique is equivalent to making measurements at $\phi = \pi/2M$. The sensitivity at $\phi = \pi/2M$ in the presence of detection noise can then be expressed as:

$$\Delta\phi^{-1}\Big|_{\phi=\frac{\pi}{2M}} \approx \frac{\sqrt{2}S^2 \sin\mu}{\sqrt{(S\sin\mu)^2 + \Delta S_{\text{DN}}^2}} \tag{26}$$

As can be seen, in the absence of detection noise, the sensitivity employing the hopping technique is the same as what we found earlier when the measurement is made at $\phi = 0$.

The robustness of a protocol against detection noise is defined as the value of $\Delta S_{\text{ES}}$ for which the sensitivity is reduced by a factor of $\sqrt{2}$. From Eq. (16), we can see that the robustness is given simply by the value of the operating point QPN [30] for any protocol. Therefore, the enhancement of the robustness compared to the protocols without spin squeezing is just the noise amplification factor (NAF), as shown in Figure 8(a3) and (b3). The robustness enhancement for the CESP and the protocols without spin squeezing is unity while the robustness for the SCSP is $\sqrt{N}$. For the GESP, in the range of the plateau, the robustness enhancement is $\sqrt{2S}\cos\mu$, which is higher than unity and increases to $\sim\sqrt{N}$ as $\mu$ approaches π/2.

# 6. Fragility of the squeezing protocols against decoherence mechanisms

In the preceding section, we have discussed the varying degrees of robustness of these protocols against detection noise. However, another factor that affects the signal to noise ratio is the reduction of the signal amplitude. As $\mu$ increases, the signal amplitude will decrease due to the experimental imperfections including cavity decay, residual spontaneous emission, and collisions with background particles. As we show later, some imperfections can reduce the signal amplitude to zero as $\mu$ approaches $\pi/2$. Therefore, despite the high robustness against the detection noise, a value of $\mu$ close to $\pi/2$ may not be the optimal choice. This is why studying the GESP, which yields maximum ideal enhancement for a broad range of values of $\mu$, is important even though the SCSP has been well investigated. In this section, we address the effects of these imperfections on the GESP.

Consider first the effect of cavity decay during the squeezing and unsqueezing step. We note first that both Ref. 1 and Ref. 11 have considered the effect of this decay carefully, and have produced results that can be extended easily to GESP, as summarized below. Briefly, the cavity decay is described by the Lindblad operator $L = \sqrt{\gamma} S_z$, where $\gamma \equiv \chi\kappa/|\delta|$, with $\chi$ being the characteristic parameter of the OATS Hamiltonian, $\kappa$ being the cavity decay rate, and $\delta$ being the probe detuning relative to the cavity resonant frequency. The effect of the cavity decay can be expressed as

$$\left\langle S_{\substack{x\\y}} \right\rangle = e^{-\gamma t/2} \left\langle S_{\substack{x\\y}} \right\rangle_{\text{ideal}} = e^{-\mu\kappa/|\delta|} \left\langle S_{\substack{x\\y}} \right\rangle_{\text{ideal}} \tag{27}$$

$$\left\langle S_{\substack{x\\y}}^2 \right\rangle = \frac{1+e^{-2\gamma t}}{2}\left\langle S_{\substack{x\\y}}^2 \right\rangle_{\text{ideal}} + \frac{1-e^{-2\gamma t}}{2}\left\langle S_{\substack{y\\x}}^2 \right\rangle_{\text{ideal}} \tag{28}$$

where the subscript "ideal" corresponds to the condition where the cavity decay is ignored. In Eq. (27), we have used the fact that $\chi t = 2\mu$ since the overall cavity-atom interaction includes both the squeezing and the unsqueezing processes. In Ref. [1], these results were considered specifically for $\mu \approx 1/\sqrt{N}$ which produces the maximum enhancement in sensitivity for the CESP. In Ref. 11 these results were considered specifically for $\mu = \pi/2$ which produces the maximum enhancement in sensitivity for the SCSP. Here, we are pointing out the fact that these apply for all values of $\mu$. As such, a consideration of these equations can be used to understand the effect of cavity decay on the sensitivity of the GESP.

As can be seen from Eq. (27), the signal amplitude equals its idea value multiplied by $\exp(-\gamma t/2)$. On the other hand, the variance of the signal varies by a factor that depends on the phase shift according to Eq. (28), since a phase shift mixes $S_x$ and $S_y$. However, the protocols are implemented at the phase shift where the variance of the measured operator, for example, $S_x$, is maximized, as discussed in Section 5, and thus the variance of $S_y$ is negligible. This fact can be proven analytically in the limits where $\mu \to 0$ and $\mu \to \pi/2$. Considering that at the phase shift where the protocols are implemented, the signal $\left\langle S_{\substack{x\\y}} \right\rangle \approx 0$, we can see from Eq. (28) that the variance of the measured operator is also reduced. Consequently, the major effect of the cavity decay is the reduction of the signal amplitude, while the variation in the noise is negligible. As can be seen from Eq. (27), increasing the magnitude of the probe detuning, $|\delta|$, will suppress the effect of the cavity decay. However, as we discuss in the next paragraph, increasing $|\delta|$ will enhance the effect of the spontaneous emission. This trade-off between the suppressions of the effects of the cavity decay and the spontaneous emission is also discussed in Refs. [1, 11].

Consider next the effect of spontaneous emission. In both Ref. [1], an analysis of the effect of cavity decay and spontaneous emission was presented for the CESP. In Ref. [11], a similar analysis was presented for the SCSP. However, both analyses are somewhat incomplete in this context. A key reason for this is that while the analyses of these protocols assume that the quantum state of the atoms always remains limited to a superposition of the $N+1$ symmetric collective states due to the nature of the coherent interactions involved, spontaneous emission can populate any of the $2^N$ collective states, which include both the symmetric and asymmetric ones [31]. Given that $2^N$ is an extremely large number for even a modest value of *N*, a comprehensive numerical analysis for the effect of spontaneous emission is nearly impossible, except for very small values of *N*. As such, in both Ref. [1] and Ref. [11] the effect of spontaneous emission has been treated heuristically, thus generating conclusions that are not very robust. One must rely on experiments to see the exact degree to which the degree of enhancement achievable under these protocols may be limited by the residual spontaneous emission. Nevertheless, a characteristic parameter related to the spontaneous emission, namely the effective decay rate of state $|\boldsymbol{z}_0\rangle$ and $|-\boldsymbol{z}_0\rangle$, can be found to be $\Gamma_{\text{eff}} = \Gamma\Omega^2/4\Delta^2 \approx \chi\Gamma|\delta|/(2g)^2$. The signal amplitude will decrease to the ideal value times $\exp(-\alpha\Gamma_{\text{eff}}t) = \exp\left(-2\alpha\mu\Gamma|\delta|/(2g)^2\right)$, where the value of $\alpha$ is in the interval $1 \le \alpha \le N$ and depends on the value of $\mu$ [1,11]. The spontaneous emission will also increase the noise level by changing the populations of, for example, $|\boldsymbol{x}_0\rangle$ and $|-\boldsymbol{x}_0\rangle$ states. However, the effective rate of the decay from state $|\boldsymbol{x}_0\rangle$ to state $|-\boldsymbol{x}_0\rangle$ is the same as the effective decay rate of the opposite direction. Therefore, for $N \gg 1$, the fluctuation of the populations of state $|\boldsymbol{x}_0\rangle$ and state $|-\boldsymbol{x}_0\rangle$ due to the spontaneous emission is negligible. Similar to the situation for the cavity decay, the major effect of the spontaneous emission is also the reduction of the signal amplitude. We can see that a direct way to suppress the effect of the spontaneous emission is to increase the single photon Rabi frequency. By maximizing the factor describing the net reduction of the signal amplitude $\exp(-\gamma t - \alpha \Gamma_{\text{eff}} t)$, we can find the optimal probe detuning to be $g\sqrt{2\kappa/\alpha\Gamma} \equiv \kappa\sqrt{C/2\alpha}$, and the maximum value of this factor to be $\exp\left(-2\mu\sqrt{2\alpha\kappa\Gamma/(2g)^2}\right) \equiv \exp\left(-2\mu\sqrt{2\alpha/C}\right)$, where $C \equiv (2g)^2/\kappa\Gamma$ is the single-atom cooperativity of the cavity. Because the increase in noise due to the cavity decay and the spontaneous emission is negligible, the reduction of the signal amplitude is just the reduction of the sensitivity due to these two imperfections. If $C \gg N$, then the effects of the cavity decay and the spontaneous emission are definitely negligible, because the upper limit of $\alpha$ is $N$. However, this requirement is nearly impossible to meet for the number of atoms typically used in an experiment. For $\mu$ close to $\pi/2$, it is very likely that $\alpha \gg C$, since for the Schrödinger cat state, a single spontaneous emission will destroy the coherence between the two CSSs constituting the Schrödinger cat state. Therefore, despite the increasing robustness against detection noise, $\mu = \pi/2$ may not be the optimal choice that will maximize the sensitivity. If it is required that the signal amplitude be reduced by a factor less than $e^2$ due to cavity decay and residual spontaneous emission, the experiment should be implemented under the condition $\mu\sqrt{2\alpha/C} \le 1$. The exact value of $\alpha$ needs to be determined with numerical simulations, which, as discussed earlier in this section, is difficult and beyond the scope of this paper. However, the value of $\alpha$ can be approximated by the PMF, since the PMF also increases from unity to $N$ as $\mu$ increases from 0 to $\pi/2$. Under this approximation, in the interval of the sensitivity plateau, the condition described above gives $\mu \leq \sqrt{C/\sqrt{2}N}$.

Finally, we consider the effect of collisions with background particles. Such collisions can produce many deleterious effects, including flipping the state of an atom, introducing an arbitrary quantum phase, and transferring momentum to an atom. In principle, such collisions can be made negligible by using ultrahigh vacuums produced under cryogenic conditions [32,33]. For example, in Refs. [34,35,36], which address this issue in the context of attempts to create the macroscopic superposition of nanoparticles, it has been shown that collisions with background particles become negligible for a vacuum of $\approx 10^{-16}$ Torr. Such pressures have been previously realized in cryogenic environments [37]. However, use of a such a vacuum may not be suitable for many practical applications of a spin-squeezed sensor. As such, in many situations, the rate of collisions with background particles may not be insignificant. Developing a comprehensive model of the collisions would depend on the specific species of atoms used and the background particles, and is beyond the scope of this paper. As such, here we simply consider one effect of collisions, as an illustrative example: ejection of an atom out of the ensemble, producing a separation large enough to exclude further interaction with the cavity mode. Although this example represents an incomprehensive description of the effect of collisions, it will yield some insight into the fact that the fragility against collisions increases as $\mu$ increases. For the ejection mechanism, we estimate the degree of reduction in the fringe contrast expected as a function of the mean number of atoms lost during collisions with background particles. For a given level of vacuum employed in the experiment, such an analysis can serve as a guideline for determining the optimal choice of operating parameters for any of the protocols discussed above.

For all the protocols employing one-axis-twist squeezing (OATS) discussed above (namely the SCSP, the GESP, and the CESP), the CSS $|\hat{\boldsymbol{x}}\rangle$ will go through the squeezing process and the inverse squeezing process. The final state will be restored to $|\hat{\boldsymbol{x}}\rangle$ in the absence of any phase shift if the atoms are not perturbed in any way between these two steps. For the SCSP and the GESP, the quantum operator measured in the end is $S_x$, with the corresponding signal being $\langle S_x \rangle$. This signal takes its maximum value of $\langle S_x \rangle = S$ in the absence of any collision between the atoms and the background particles. Next, we calculate $\langle S_x \rangle$ in the presence of such collisions. To develop the mathematical model to account for the effect of such collisions, we first define the spin operators for the atoms that have collided with background particles and the atoms that have not. We assume without loss of generality that the first $\breve{N}$ atoms have collied with background particles. Then the total spin operator for these atoms is denoted as $\breve{\boldsymbol{S}} \equiv \sum_{j=1}^{\breve{N}} \boldsymbol{s}_j$. The spin operator for the rest of the atoms is denoted as $\widehat{\boldsymbol{S}} \equiv \sum_{j=\breve{N}+1}^{N} \boldsymbol{s}_j$. It is obvious that collisions that happen before the squeezing process or after the inverse squeezing process simply reduces the effective value of $N$. As such, these types of collisions are negligible if $\breve{N} \ll N$. Therefore, we only consider the case where the collisions happen between the squeezing and the inverse squeezing process.

Under this assumption, the effect of a collision is modeled as an event under which an atom that has collided with a background particle follows a kinetic trajectory that eludes the inverse squeezing process. Mathematically, the inverse squeezing process is expressed as $\exp\left(\mathrm{i}\mu \widehat{S}_z^2\right)$ in the presence of the collisions. Therefore, the final state can be expressed as

$$\left|\psi\right\rangle = \mathrm{e}^{\mathrm{i}\mu \breve{S}_z^2} \mathrm{e}^{-\mathrm{i}\mu S_z^2} \left|\hat{\boldsymbol{x}}\right\rangle \tag{29}$$

The quantum operator we measure in the end is $\breve{S}_x$ because any atom that has not undergone the inverse squeezing process will not be detected. As shown in detail in Appendix C, the corresponding signal is calculated to be

$$\left\langle\psi\right|\breve{S}_x\left|\psi\right\rangle = \left(S - \breve{S}\right)\cos^{2\breve{S}}\mu \tag{30}$$

where $\breve{S} = \breve{N}/2$ is the total spin of the atoms that have undergone collisions. For $\mu = 0$, the signal is reduced to $\left(S - \breve{S}\right)$, as expected, corresponding to simple exclusion of the atoms involved in collisions. For $\mu = \pi/2$, which corresponds to the Schrödinger cat state, the signal is zero, for $\breve{N} \geq 1$. Accordingly, the Schrödinger cat state is the most fragile to collisions. Thus, if one were to employ the SCSP experimentally, the vacuum level must be such that the probability of even one collision per cycle of the squeezing-unsqueezing sequence is significantly less than unity. To evaluate quantitatively the fragility of a protocol, we define the maximum tolerable value of $\breve{N}$ as the number of atoms involved in collisions that will reduce the signal contrast by a factor of $e$ compared to the $\mu = 0$ case. This maximum tolerable $\breve{N}$ is calculated to be $\left(-1/\ln\left(\cos\mu\right)\right)$. In the limit $\breve{N} \gg 1$, the maximum tolerable $\breve{N}$ is thus approximately $2/\mu^2$, meaning that, in order maintain significant fringe visibility, the experiment needs to be implemented under the condition $\mu \leq 1/\sqrt{\breve{S}}$.

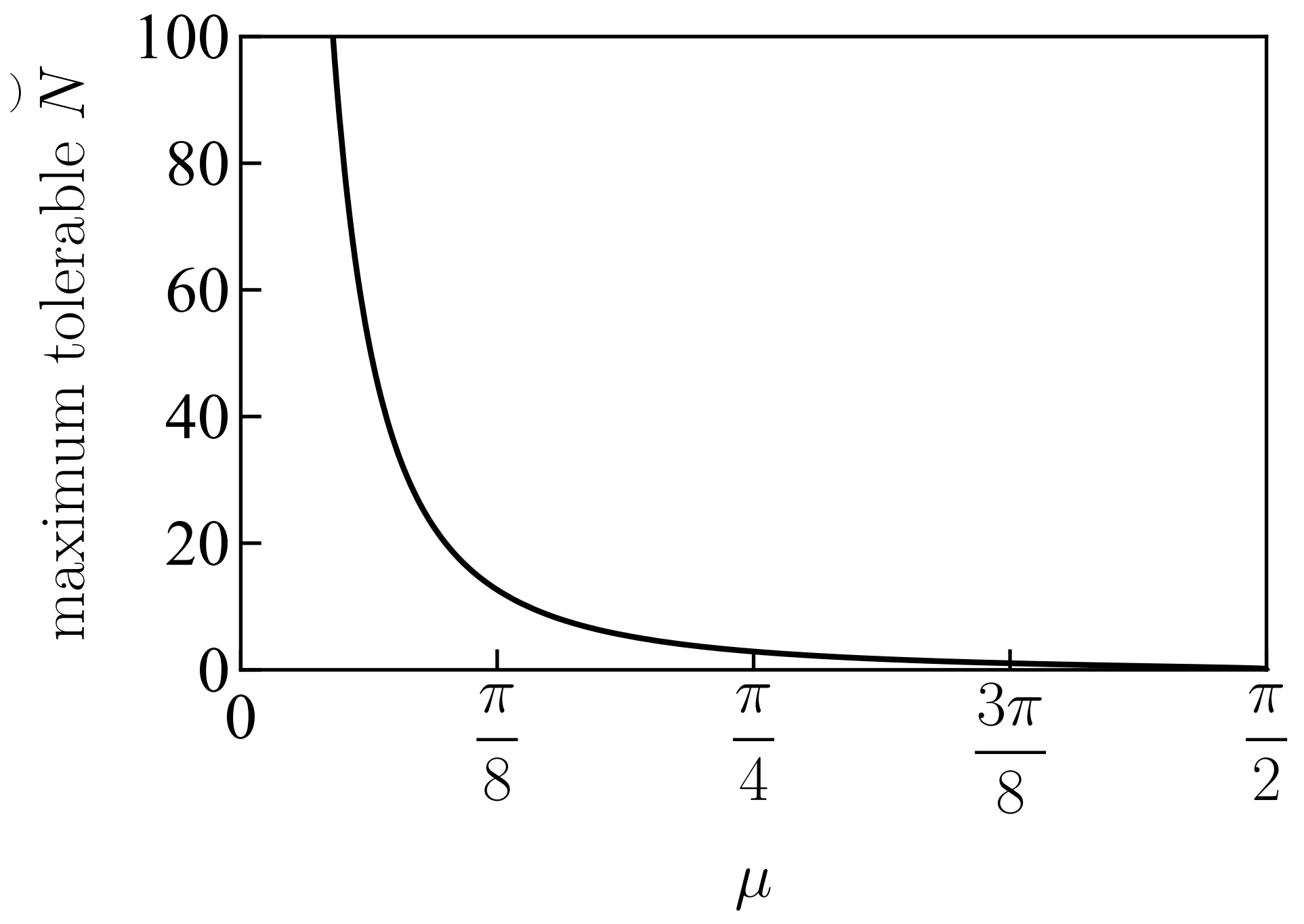

Figure 13 Maximum tolerable $\breve{N}$, defined as the number of atoms involved in collisions that will reduce the signal contrast by a factor of e compared to the $\mu = 0$ case. This maximum tolerable $\breve{N}$ is calculated to be $\left(-1/\ln(\cos\mu)\right)$. In the limit $\breve{N} \gg 1$, the maximum tolerable $\breve{N}$ is approximately $2/\mu^2$.

As noted earlier, an apparatus employing a cryogenic vacuum at the level of $\sim 10^{-16}$ Torr may be suitable for implementing the SCSP, satisfying the condition that the probability of even one collision between the squeezing and unsqueezing processes in each cycle is significantly less than unity. For a vacuum system that can ensure a significant proportion of the experiment cycles that are not affected by the background particles, the SCSP can still approach the Heisenberg limit because averaging over the affected and the unaffected measurement results will only reduce the signal contrast. For a vacuum system where the number of expected collisions in each cycle is far greater than unity, one needs to ensure that the condition mentioned above (i.e., $\mu \le 1/\sqrt{S}$ ) is satisfied. Thus, the optimal choice of the squeezing parameter for the GESP would be determined by a balance between the reduction of the signal amplitude and the degree of robustness against detection noise.

# 7. Conclusion

In this paper, we describe the advantages of the Generalized Echo Squeezing Protocol (GESP) over the Conventional Echo Squeezing Protocol (CESP). We point out that there are two different versions of the GESP, denoted as GESP-e and GESP-o, each optimized for one of the two possible parities (even or odd) of the total number of atoms. We present the analytical expressions for the sensitivity of both the GESP-e and the GESP-o. The GESP is a generalization of the Schrödinger Cat State Protocol (SCSP) with the value of $\mu$ being an arbitrary number rather than π/2. We find that in the interval $4S^{-1/2} \leq \mu \leq \left(\pi/2 - S^{-1/2}\right)$, the sensitivity of the GESP is approximately the Heisenberg limit within a factor of $\sqrt{2}$. For a large *N*, this plateau interval is almost the whole range from 0 to π/2. Therefore, it is possible to operate a sensor over a wide interval of $\mu$ without changing the sensitivity. On the contrary, the CESP only works for a very small interval of $\mu$. The sensitivity of CESP drops to $\sim 0$ once the value of $\mu$ becomes greater than $2S^{-1/2}$, which is a very small number for a large *N*. We also show that, in contrast to the CESP, the sensitivity of the GESP is close to the quantum Cramér-Rao bound over the whole range of $\mu$, indicating that the information of the phase shift contained in the quantum state is near-optimally extracted for any value of $\mu$. In the interval $4S^{-1/2} \leq \mu \leq \left(\pi/2 - S^{-1/2}\right)$ and for $S \gg 1$, the central fringe of the signal for the GESP does not depend on the parity of *N* as well as whether we use GESP-e or GESP-o. We also show how the enhancement in sensitivity in the case of the GESP is due to a combination of two parameters: the phase magnification factor (PMF) and the noise amplification factor (NAF). As the value of $\mu$ increases, both PMF and NAF increase, while keeping the ratio of PMF/NAF essentially constant, yielding a net enhancement of sensitivity that is a factor of $\sqrt{2}$ lower than the Heisenberg limit over the whole plateau interval. An important consequence of this behavior is that the robustness of the GESP against detection noise easily exceeds that of the CESP for a broad range of values of $\mu$. As such, in the context of an experimental study, it should be possible to achieve a net enhancement of sensitivity higher than that for the CESP, under typical conditions where detection noise exceeds the quantum projection noise of an unsqueezed ensemble with the same number of atoms. Finally, we have analyzed the fragility of the GESP against decoherence mechanisms, and show how a balance between the fragility against decoherence mechanisms and the robustness against detection noise would in practice determine the optimal choice of parameters for the GESP.

## Acknowledgement:

This work has been supported equally in parts by the Department of Defense Center of Excellence in Advanced Quantum Sensing under Army Research Office grant number W911NF202076, and the U.S. Department of Energy, Office of Science, National Quantum Information Science Research Centers, Superconducting Quantum Materials and Systems Center (SQMS) under contract number DE-AC02-07CH11359.

# Appendix A

We first consider an ensemble with $N$ atoms initially in state $|\hat{\boldsymbol{x}}\rangle$, where $N$ is even. After the ensemble undergoes the squeezing process with $\mu = \pi/2$, the final state is an *x*-oriented Schrödinger cat state $\left[|\hat{\boldsymbol{x}}\rangle + (-1)^{N/2}\, i\,|-\hat{\boldsymbol{x}}\rangle\right]\big/\sqrt{2}$. If an atom in state $|\pm\hat{\boldsymbol{z}}_0\rangle$ is added to the ensemble before the squeezing process, the final state after the squeezing process can be expressed as $\exp\left(-i(\pi/2)(S_z + s_z)^2\right)|\hat{\boldsymbol{x}}\rangle \otimes |\pm\hat{\boldsymbol{z}}_0\rangle$, where $S_z$ is the spin operator acting on the original $N$ atoms and $s_z$ is the spin operator acting on the additional atom. The squeezing propagator for the $N+1$ atoms can be written as a product of three terms:

$$e^{-i(\pi/2)(S_z+s_z)^2} = e^{-i(\pi/2)s_z^2} e^{-i\pi s_z S_z} e^{-i(\pi/2)S_z^2} = e^{-i(\pi/2)(1/4)} e^{-i\pi s_z S_z} e^{-i(\pi/2)S_z^2} \tag{31}$$

The first term is simply a common phase factor since $s_z^2 = 1/4$ regardless of the state of the additional atom. The second term represents an effective interaction between the additional atom and the original $N$ atoms, resulting from the non-linearity of the squeezing process. With the additional atom in state $|\pm\hat{\boldsymbol{z}}_0\rangle$, this term becomes a rotation operator for the original $N$ atoms, with the rotation angle being $\pm\pi/2$, because $\exp(-i\pi s_z S_z)|\pm\hat{\boldsymbol{z}}_0\rangle = \exp\left(-i\pi(\pm 1/2)S_z\right)|\pm\hat{\boldsymbol{z}}_0\rangle$. The third term is the squeezing operator for the original $N$ atoms. Consequently, the final state can be expressed (separately for the different initial states of the extra atom for clarity) as

$$\begin{aligned}\left(e^{-i(\pi/2)S_z} e^{-i(\pi/2)S_z^2}|\hat{\boldsymbol{x}}\rangle\right)\otimes|\hat{\boldsymbol{z}}_0\rangle &= \left\{e^{-i(\pi/2)S_z}\frac{1}{\sqrt{2}}\left[|\hat{\boldsymbol{x}}\rangle + (-1)^{N/2}\, i\,|-\hat{\boldsymbol{x}}\rangle\right]\right\}\otimes|\hat{\boldsymbol{z}}_0\rangle \\ &= \frac{1}{\sqrt{2}} i^{-N/2}\left[|\hat{\boldsymbol{y}}\rangle + (-1)^{N/2}\, i\,|-\hat{\boldsymbol{y}}\rangle\right]\otimes|\hat{\boldsymbol{z}}_0\rangle\end{aligned} \tag{32}$$

$$\left(e^{-i(-\pi/2)S_z}e^{-i(\pi/2)S_z^2}\,|\hat{\boldsymbol{x}}\rangle\right)\otimes|-\hat{\boldsymbol{z}}_0\rangle=\left\{e^{-i(-\pi/2)S_z}\,\frac{1}{\sqrt{2}}\Big[|\hat{\boldsymbol{x}}\rangle+(-1)^{N/2}\,i|-\hat{\boldsymbol{x}}\rangle\Big]\right\}\otimes|-\hat{\boldsymbol{z}}_0\rangle$$
$$=\frac{1}{\sqrt{2}}\,i^{-N/2+1}\Big[|\hat{\boldsymbol{y}}\rangle-(-1)^{N/2}\,i|-\hat{\boldsymbol{y}}\rangle\Big]\otimes|-\hat{\boldsymbol{z}}_0\rangle \tag{33}$$

The state of the original $N$ atoms shown in Eqs. (32) and (33) is the *x*-directed Schrödinger cat state rotated by $\pm\pi/2$, which is the *y*-directed Schrödinger cat state $\Big[|\hat{\boldsymbol{y}}\rangle\pm(-1)^{N/2}\,i|-\hat{\boldsymbol{y}}\rangle\Big]\Big/\sqrt{2}$. This show that the addition of a single atom results in a rotation of the Schrödinger cat for the original $N$ atoms around the *z* axis by $\pm\pi/2$, depending on the initial state of the extra atom (i.e., $|\pm\hat{\boldsymbol{z}}_0\rangle$)

As such, if the additional atom is initially in state $|\hat{\boldsymbol{x}}_0\rangle=\big(|\hat{\boldsymbol{z}}_0\rangle+|-\hat{\boldsymbol{z}}_0\rangle\big)\big/\sqrt{2}$, the final state will be a superposition of $i^{-N/2}\Big[|\hat{\boldsymbol{y}}\rangle+(-1)^{N/2}\,i|-\hat{\boldsymbol{y}}\rangle\Big]\Big/\sqrt{2}\otimes|\hat{\boldsymbol{z}}_0\rangle$ and $i^{-N/2+1}\Big[|\hat{\boldsymbol{y}}\rangle-(-1)^{N/2}\,i|-\hat{\boldsymbol{y}}\rangle\Big]\Big/\sqrt{2}\otimes|-\hat{\boldsymbol{z}}_0\rangle$. Noting that $|\pm\hat{\boldsymbol{z}}_0\rangle=\big(|\hat{\boldsymbol{y}}_0\rangle\pm i|-\hat{\boldsymbol{y}}_0\rangle\big)\big/\sqrt{2}$, we can see that the final state is $i^{-N/2}\big(|\hat{\boldsymbol{y}}\rangle\otimes|\hat{\boldsymbol{y}}_0\rangle+i|-\hat{\boldsymbol{y}}\rangle\otimes|-\hat{\boldsymbol{y}}_0\rangle\big)\big/\sqrt{2}$, which is a *y*-directed Schrödinger cat state of the $N+1$ atoms.

To summarize, an addition of a single atom in state $|\pm\hat{\boldsymbol{z}}_0\rangle$ results in a rotation around the *z* axis by an angle of $\pm\pi/2$ and thus rotates an *x*-directed Schrödinger cat state (for example) to a *y*-directed Schrödinger cat state. If the additional atom is initially in state $|\hat{\boldsymbol{x}}_0\rangle$, the final state is a superposition of two *y*-directed Schrödinger cat states entangled to the additional atom. This superposition can be rewritten as a *y*-directed Schrödinger cat state of the whole ensemble including the additional atom.

# Appendix B

To find the explicit expression of the signal shown in Eq. (8), we need to calculate $\left\langle \tilde{S}^2_{\substack{x\\y}} \right\rangle$ and $\left\langle \tilde{S}_{\substack{x\\y}} S_x \tilde{S}_{\substack{x\\y}} \right\rangle$. With the relations [3] $S_{\substack{x\\y}} = S_+ \pm \mathrm{i} S_y$, $[S_z, S_\pm] = \pm S_\pm$, $[S_+, S_-] = 2S_z$, and $f(S_z) S_\pm = S_\pm f(S_z \pm 1)$, we can rearrange the terms in the $\left\langle \tilde{S}_{\substack{x\\y}} S_x \tilde{S}_{\substack{x\\y}} \right\rangle$ in the normal order or anti-normal order, as needed. Then we can do the calculation using the generating functions [26], and the two terms mentioned above are calculated to be

$$\left\langle \tilde{S}^2_{\substack{x\\y}} \right\rangle = \frac{1}{2} S \left( S + \frac{1}{2} \right) \pm \frac{1}{2} S \left( S - \frac{1}{2} \right) \cos^{2S-2} 2\mu \tag{34}$$

$$\left\langle \tilde{S}_{\substack{x\\y}} S_x \tilde{S}_{\substack{x\\y}} \right\rangle = \frac{1}{2} S \left( S - \frac{1}{2} \right)(S+1) \cos 2\mu \pm \frac{1}{2} S \left( S - \frac{1}{2} \right)(S-1) \cos^{2S-3} 2\mu \pm \frac{1}{2} S^2 \cos^{2S-1} 2\mu \tag{35}$$

Substituting Eqs. (34) and (35) into Eq. (8), we obtain the expression of the signal shown in Eq. (9), with the vectors in the expression given by

$$\boldsymbol{a}_{10} = \frac{1}{2} S \begin{bmatrix} -S\left( S + \frac{1}{2} \right) \\ \left( S - \frac{1}{2} \right)(S+1) \\ 0 \end{bmatrix}, \qquad \boldsymbol{a}_{11} = \frac{1}{2} S \begin{bmatrix} 0 \\ \left( S - \frac{1}{2} \right)(S-1) \\ -S\left( S - \frac{1}{2} \right) \\ S \end{bmatrix} \tag{36}$$

and

$$\boldsymbol{b}_0 = \begin{bmatrix} 1 \\ \cos 2\mu \\ \cos^2 2\mu \end{bmatrix}, \qquad \boldsymbol{b}_1 = \cos^{2S-4} 2\mu \begin{bmatrix} 1 \\ \cos 2\mu \\ \cos^2 2\mu \\ \cos^3 2\mu \end{bmatrix} \tag{37}$$

To calculate the variance of $S_x$, we also need to calculate $\left\langle \psi_{\substack{e\\o}} \middle| S_x^2 \middle| \psi_{\substack{e\\o}} \right\rangle$. Similar to Eq. (8), this expectation value can be expressed as

$$\left\langle \psi_{\substack{e\\o}} \middle| S_x^2 \middle| \psi_{\substack{e\\o}} \right\rangle = S^2 + \phi^2 \left( \left\langle \tilde{S}_{\substack{x\\y}} S_x^2 \tilde{S}_{\substack{x\\y}} \right\rangle - S^2 \left\langle \tilde{S}_{\substack{x\\y}}^2 \right\rangle \right) + \mathcal{O}\left(\phi^4\right) \tag{38}$$

The first term in the parenthesis in Eq. (38) is calculated to be

$$\begin{aligned} \left\langle \tilde{S}_{\substack{x\\y}} S_x^2 \tilde{S}_{\substack{x\\y}} \right\rangle &= \frac{1}{8} S\left(2S^2 + 3S - 1\right) + \frac{1}{2} S\left(S - \frac{1}{2}\right)\left(S-1\right)\left(S + \frac{3}{2}\right)\cos^2 2\mu \\ &\pm \frac{1}{2} S\left(S - \frac{1}{2}\right)\left(S-1\right)\left(S - \frac{3}{2}\right)\cos^{2S-4} 2\mu \pm \frac{1}{4} S\left(S - \frac{1}{2}\right)\left(5S-1\right)\cos^{2S-2} 2\mu \end{aligned} \tag{39}$$

The second term in the parenthesis in Eq. (38) is determined by Eq. (34). Substituting Eqs. (34) and (39) in to Eq. (38), we have

$$\left\langle \psi_{\substack{e\\o}} \middle| S_x^2 \middle| \psi_{\substack{e\\o}} \right\rangle = S^2 + \phi^2 \left( \boldsymbol{a}_{20} \cdot \boldsymbol{b}_0 \pm \boldsymbol{a}_{21} \cdot \boldsymbol{b}_1 \right) + \mathcal{O}\left(\phi^4\right) \tag{40}$$

where

$$\boldsymbol{a}_{20} = \frac{1}{2} S\left(S - \frac{1}{2}\right) \begin{bmatrix} -\left(S - \frac{1}{2}\right)\left(S+1\right) \\ 0 \\ \left(S-1\right)\left(S + \frac{3}{2}\right) \end{bmatrix}, \qquad \boldsymbol{a}_{21} = \frac{1}{2} S\left(S - \frac{1}{2}\right) \begin{bmatrix} \left(S-1\right)\left(S - \frac{3}{2}\right) \\ 0 \\ -\left(S^2 - \frac{5}{2}S + \frac{1}{2}\right) \\ 0 \end{bmatrix} \tag{41}$$

Then the variance is calculated to be the result shown in Eq. (10), with the vectors in the expression given by

$$\boldsymbol{a}_{30} = \frac{1}{2}S\begin{bmatrix} S^3 + S^2 + \frac{3}{4}S - \frac{1}{4} \\ -2S\left(S+1\right)\left(S-\frac{1}{2}\right) \\ \left(S-\frac{1}{2}\right)\left(S-1\right)\left(S+\frac{3}{2}\right) \end{bmatrix}, \qquad \boldsymbol{a}_{31} = \frac{1}{2}S\begin{bmatrix} \left(S-\frac{1}{2}\right)\left(S-1\right)\left(S-\frac{3}{2}\right) \\ -2S\left(S-\frac{1}{2}\right)\left(S-1\right) \\ \left(S-\frac{1}{2}\right)\left(S^2+\frac{5}{2}S-\frac{1}{2}\right) \\ -2S^2 \end{bmatrix} \tag{42}$$

# Appendix C

Noting that $\boldsymbol{S} = \widehat{\boldsymbol{S}} + \breve{\boldsymbol{S}}$ and $\left[\widehat{\boldsymbol{S}}, \breve{\boldsymbol{S}}\right] = 0$, the final state given by Eq. (29) can be further written as

$$\left|\psi\right\rangle = \mathrm{e}^{\mathrm{i}\mu\widehat{S}_z^2}\mathrm{e}^{-\mathrm{i}\mu\left(\widehat{S}_z+\breve{S}_z\right)^2}\left|\hat{\boldsymbol{x}}\right\rangle = \mathrm{e}^{-\mathrm{i}\mu\breve{S}_z^2}\mathrm{e}^{-\mathrm{i}2\mu\widehat{S}_z\breve{S}_z}\left|\hat{\boldsymbol{x}}\right\rangle \tag{43}$$

The signal in the presence of collisions between the atoms and the background particles can then be expressed as

$$\begin{aligned} \left\langle\psi\right|\widehat{S}_x\left|\psi\right\rangle &= \left\langle\hat{\boldsymbol{x}}\right|\mathrm{e}^{\mathrm{i}2\mu\widehat{S}_z\breve{S}_z}\mathrm{e}^{\mathrm{i}\mu\breve{S}_z^2}\widehat{S}_x\mathrm{e}^{-\mathrm{i}\mu\breve{S}_z^2}\mathrm{e}^{-\mathrm{i}2\mu\widehat{S}_z\breve{S}_z}\left|\hat{\boldsymbol{x}}\right\rangle = \left\langle\hat{\boldsymbol{x}}\right|\mathrm{e}^{\mathrm{i}2\mu\widehat{S}_z\breve{S}_z}\widehat{S}_x\mathrm{e}^{-\mathrm{i}2\mu\widehat{S}_z\breve{S}_z}\left|\hat{\boldsymbol{x}}\right\rangle \\ &= \prod_{j=1}^{\breve{N}} {}_j\left\langle\hat{\boldsymbol{x}}_0\right|\mathrm{e}^{\mathrm{i}2\mu s_{jz}\widehat{S}_z}\widehat{S}_x\mathrm{e}^{-\mathrm{i}2\mu s_{jz}\widehat{S}_z}\left|\hat{\boldsymbol{x}}_0\right\rangle_j \\ &= \left[\left\langle\hat{\boldsymbol{x}}_0\right|\mathrm{e}^{\mathrm{i}2\mu s_{1z}\widehat{S}_z}\widehat{S}_x\mathrm{e}^{-\mathrm{i}2\mu s_{1z}\widehat{S}_z}\left|\hat{\boldsymbol{x}}_0\right\rangle\right]^{\breve{N}} \end{aligned} \tag{44}$$

The step in the first line of Eq. (44) still makes use of the commutation relation $\left[\widehat{\boldsymbol{S}}, \breve{\boldsymbol{S}}\right] = 0$. The step in second line of Eq. (44) is based on the definition $\breve{\boldsymbol{S}} \equiv \sum_{j=1}^{\breve{N}} \boldsymbol{s}_j$. The step in the third line of Eq.

(44) is based on the fact that the first $\breve{N}$ atoms are interchangeable. Considering that $|\hat{\boldsymbol{x}}_0\rangle = \left(|\hat{\boldsymbol{z}}_0\rangle + |-\hat{\boldsymbol{z}}_0\rangle\right)/\sqrt{2}$ and $s_z|\pm\hat{\boldsymbol{z}}_0\rangle = \pm 1/2|\pm\hat{\boldsymbol{z}}_0\rangle$, Eq. (44) can be further written as

$$\begin{aligned}\langle\psi|\hat{S}_x|\psi\rangle &= \frac{1}{2}\left[\langle\hat{\boldsymbol{z}}_0|\hat{\boldsymbol{z}}_0\rangle\ {}_{\mathrm{c}}\langle\hat{\boldsymbol{x}}|\mathrm{e}^{\mathrm{i}\mu\hat{S}_z}\hat{S}_x\mathrm{e}^{-\mathrm{i}\mu\hat{S}_z}|\hat{\boldsymbol{x}}\rangle_{\mathrm{c}}\right]^{\breve{N}} + \frac{1}{2}\left[\langle-\hat{\boldsymbol{z}}_0|-\hat{\boldsymbol{z}}_0\rangle\ {}_{\mathrm{c}}\langle\hat{\boldsymbol{x}}|\mathrm{e}^{\mathrm{i}\mu\hat{S}_z}\hat{S}_x\mathrm{e}^{-\mathrm{i}\mu\hat{S}_z}|\hat{\boldsymbol{x}}\rangle_{\mathrm{c}}\right]^{\breve{N}} \\ &= \left[{}_{\mathrm{c}}\langle\hat{\boldsymbol{x}}|\mathrm{e}^{\mathrm{i}\mu\hat{S}_z}\hat{S}_x\mathrm{e}^{-\mathrm{i}\mu\hat{S}_z}|\hat{\boldsymbol{x}}\rangle_{\mathrm{c}}\right]^{\breve{N}} = \left(S-\breve{S}\right)\cos^{2\breve{S}}\mu\end{aligned} \tag{45}$$

where $|\hat{\boldsymbol{x}}\rangle_{\mathrm{c}} \equiv \sum_{j=1}^{\breve{N}}|\hat{\boldsymbol{x}}_0\rangle_j$ is a CSS of the atoms involved in collisions.

# References:

[1] Davis, E., Bentsen, G., & Schleier-Smith, M. (2016). Approaching the Heisenberg limit without single-particle detection. Physical review letters, 116(5), 053601.

[2] Hosten, O., Krishnakumar, R., Engelsen, N. J., & Kasevich, M. A. (2016). Quantum phase magnification. Science, 352(6293), 1552-1555.

[3] Kitagawa, M., & Ueda, M. (1993). Squeezed spin states. Physical Review A, 47(6), 5138.

[4] Sørensen, A. S., & Mølmer, K. (2002). Entangling atoms in bad cavities. Physical Review A, 66(2), 022314.

[5] Britton, J. W., Sawyer, B. C., Keith, A. C., Wang, C. C. J., Freericks, J. K., Uys, H., ... & Bollinger, J. J. (2012). Engineered two-dimensional Ising interactions in a trapped-ion quantum simulator with hundreds of spins. Nature, 484(7395), 489-492.

[6] Sørensen, A., Duan, L. M., Cirac, J. I., & Zoller, P. (2001). Many-particle entanglement with Bose–Einstein condensates. Nature, 409(6816), 63-66.

[7] Schleier-Smith, M. H., Leroux, I. D., & Vuletić, V. (2010). Squeezing the collective spin of a dilute atomic ensemble by cavity feedback. Physical Review A, 81(2), 021804.

[8] Leroux, I. D., Schleier-Smith, M. H., & Vuletić, V. (2010). Implementation of cavity squeezing of a collective atomic spin. Physical Review Letters, 104(7), 073602.

[9] Zhang, Y. L., Zou, C. L., Zou, X. B., Jiang, L., & Guo, G. C. (2015). Detuning-enhanced cavity spin squeezing. Physical Review A, 91(3), 033625.

[10] It can be shown that the description we present regarding the CESP protocol applied to a Ramsey clock is equivalent to the description of the process presented in Ref. [1], for example.

[11] Fang, R., Sarkar, R., & Shahriar, S. M. (2020). Enhancing the sensitivity of an atom interferometer to the Heisenberg limit using increased quantum noise. JOSA B, 37(7), 1974-1986. Expanded version of this paper with additional details can be found at https://arxiv.org/abs/1707.08260

[12] Haine, S. A. (2018). Using interaction-based readouts to approach the ultimate limit of detection-noise robustness for quantum-enhanced metrology in collective spin systems. Physical Review A, 98(3), 030303.

[13] Sarkar, R., Kim, M. E., Fang, R., & Shahriar, S. M. (2015). N-atom collective-state atomic interferometer with ultrahigh Compton frequency and ultrashort de Broglie wavelength, with root-N reduction in fringe width. Physical Review A, 92(6), 063612

[14] Sørensen, A., & Mølmer, K. (2000). Entanglement and quantum computation with ions in thermal motion. Physical Review A, 62(2), 022311.

[15] Leibfried, D., Barrett, M. D., Schaetz, T., Britton, J., Chiaverini, J., Itano, W. M., ... & Wineland, D. J. (2004). Toward Heisenberg-limited spectroscopy with multiparticle entangled states. Science, 304(5676), 1476-1478.

[16] Leibfried, D., Knill, E., Seidelin, S., Britton, J., Blakestad, R. B., Chiaverini, J., ... & Wineland, D. J. (2005). Creation of a six-atom 'Schrödinger cat'state. Nature, 438(7068), 639-642.
[17] Leibfried, D., & Wineland, D. J. (2018). Efficient eigenvalue determination for arbitrary Pauli products based on generalized spin-spin interactions. Journal of Modern Optics, 65(5-6), 774-779.
[18] Monz, T., Schindler, P., Barreiro, J. T., Chwalla, M., Nigg, D., Coish, W. A., ... & Blatt, R. (2011). 14-qubit entanglement: Creation and coherence. Physical Review Letters, 106(13), 130506.
[19] This magnification factor occurs only for the optimal choice of the squeezing parameter, i.e., $\mu = 1/\sqrt{N}$. When the squeezing parameter is much smaller than this, it can be shown [1,2] that the magnification factor is $\sim N\mu$.
[20] In principle, the rotation can also be measure by applying a π/2 rotation around the y-axis, followed by a measurement of $S_z$, and that is exactly what we do for the GESP-o. To see the distinction between these two approaches, it is instructive to consider the case where all four additional pulses are absent. In that case, use of a π/2 rotation around the y-axis produces a signal that is of the form $(N/2)\cos\phi$. In contrast, use of a π/2 rotation around the x-axis produces a signal that is of the form $(N/2)\sin\phi$.
[21] Bordé, C. J. (1989). Atomic interferometry with internal state labelling. Physics letters A, 140(1-2), 10-12.
[22] Kasevich, M., & Chu, S. (1991). Atomic interferometry using stimulated Raman transitions. Physical review letters, 67(2), 181.
[23] Barrett, B., Geiger, R., Dutta, I., Meunier, M., Canuel, B., Gauguet, A., ... & Landragin, A. (2014). The Sagnac effect: 20 years of development in matter-wave interferometry. Comptes Rendus Physique, 15(10), 875-883.
[24] Sarkar, R., Fang, R., & Shahriar, S. M. (2018). High-Compton-frequency, parity-independent, mesoscopic Schrödinger-cat-state atom interferometer with Heisenberg-limited sensitivity. Physical Review A, 98(1), 013636.
[25] Dicke, R. H. (1954). Coherence in spontaneous radiation processes. Physical review, 93(1), 99.
[26] Arecchi, F. T., Courtens, E., Gilmore, R., & Thomas, H. (1972). Atomic coherent states in quantum optics. Physical Review A, 6(6), 2211.
[27] Wikipedia contributors. (2021, May 10). Quantum Fisher information. In Wikipedia, The Free Encyclopedia. Retrieved 19:35, October 10, 2021, from https://en.wikipedia.org/w/index.php?title=Quantum_Fisher_information&oldid=1022484528
[28] As we have shown in Ref. [11], the values of the phase magnification factor and the noise amplification factor for the parity-matched SCSPs actually depend critically on what type of detection process is employed. Specifically, if the process of collective state detection is employed, then the results are very different from what we get if the conventional detection method is employed. Here, we are assuming that the detection process is conventional, meaning that the signal is $\langle S_x \rangle$ and the noise is $\Delta S_x$, in order to establish proper comparisons with CESP and GESP.
[29] Itano, W. M., Bergquist, J. C., Bollinger, J. J., Gilligan, J. M., Heinzen, D. J., Moore, F. L., ... & Wineland, D. J. (1993). Quantum projection noise: Population fluctuations in two-level systems. Physical Review A, 47(5), 3554.
[30] If the signal is co-sinusoidal with respect to the phase shift, then the sensitivity is independent of the value of the phase at which the measurement is made. If the hopping method is used in such a case, the sensitivity again is independent of the swing range used. However, when optimized a system for maximum robustness, it is necessary to ensure that the quantum projection noise is maximum at the swing points. We have assumed this to be the case in determining the degree of robustness for such a protocol. In contrast, if the signal is anti-symmetric with respect to the phase shift (such as in the case of the CESP), the hopping technique cannot be used. For such a case, we have assumed the operating point to be at zero phase shift, which yields the maximum sensitivity, and the robustness is evaluated at this point.
[31] Resham Sarkar, May E. Kim, Renpeng Fang, Yanfei Tu and Selim M. Shahriar, "Effects of non-idealities and quantization of the center of mass motion on symmetric and asymmetric collective states in a collective state atomic Interferometer," J. Mod. Opt. Vol. 62, Issue 15, 1253-1263 (2015).
[32] Willems, P. A., & Libbrecht, K. G. (1995). Creating long-lived neutral-atom traps in a cryogenic environment. Physical Review A, 51(2), 1403.
[33] Andresen, G. B., Ashkezari, M. D., Baquero-Ruiz, M., Bertsche, W., Butler, E., Cesar, C. L., ... & ALPHA Collaboration. (2011). Confinement of antihydrogen for 1000 seconds. arXiv preprint arXiv:1104.4982.
[34] Romero-Isart, O. (2011). Quantum superposition of massive objects and collapse models. Physical Review A, 84(5), 052121.

[35] Romero-Isart, O. (2017). Coherent inflation for large quantum superpositions of levitated microspheres. New Journal of Physics, 19(12), 123029.
[36] Pino, H., Prat-Camps, J., Sinha, K., Venkatesh, B. P., & Romero-Isart, O. (2018). On-chip quantum interference of a superconducting microsphere. Quantum Science and Technology, 3(2), 025001.
[37] Gabrielse, G., Fei, X., Orozco, L. A., Tjoelker, R. L., Haas, J., Kalinowsky, H., ... & Kells, W. (1990). Thousandfold improvement in the measured antiproton mass. Physical review letters, 65(11), 1317.